\documentclass[10pt,twocolumn,journal]{IEEEtran}
\usepackage{amssymb,amsfonts}
\usepackage[cmex10]{amsmath}
\usepackage{bbm}
\usepackage{mathtools}
\usepackage{enumerate}
\usepackage{graphicx}
\usepackage{cite}

\usepackage{caption,subcaption}

\usepackage{color}

\newcommand{\diag}{{\rm diag\;}}
\def\la{\lambda}
\def\La{\Lambda}

\def\ve{\varepsilon}

\def\r{\mathbb R}

\def\G{\mathcal G}

\def\diag{\mathop{diag}\nolimits}

\def\E{\mathbb{E}}
\def\vec{\mathop{\rm vec}\nolimits}

\def\be{\beta}

\def\be{\begin{equation}}
\def\ee{\end{equation}}
\def\ben{\begin{equation*}}
\def\een{\end{equation*}}

\newtheorem{thm}{\bf Theorem}

\newtheorem{definition}{\bf Definition}
\newtheorem{lemma}{\bf Lemma}
\newtheorem{cor}{\bf Corollary}

\newtheorem{remark}{\bf Remark}

\newtheorem{example}{\bf Example}

\def\diag{\mathop{\rm diag}\nolimits}
\def\E{\mathbb{E}}

\title{Novel Multidimensional Models of Opinion Dynamics in Social Networks}

\author{Sergey E. Parsegov, Anton~V.~Proskurnikov,~\IEEEmembership{Member,~IEEE}, Roberto Tempo,~\IEEEmembership{Fellow,~IEEE} and Noah E. Friedkin
\thanks{S.E. Parsegov is with V.A. Trapeznikov Institute for Control Sciences of Russian Academy of Sciences (ICS RAS), Moscow, Russia, {e-mail: \tt s.e.parsegov@gmail.com}}%
\thanks{A.V. Proskurnikov is with Engineering and Technology Institute (ENTEG) at the University of Groningen, The Netherlands,
and also with Institute for Problems of Mechanical Engineering of Russian Academy of Sciences (IPME RAS) and ITMO University, St. Petersburg, Russia, {e-mail: \tt anton.p.1982@ieee.org}}%
\thanks{N.E. Friedkin is with the University of Santa-Barbara, CA, USA {e-mail: \tt friedkin@soc.ucsb.edu}}%
\thanks{R. Tempo is with CNR-IEIIT, Politecnico di Torino, Italy, {e-mail: \tt roberto.tempo@polito.it }}%
\thanks{Partial funding is provided by the European Research Council (grant ERCStG-307207), CNR International Joint Lab COOPS, RFBR
(grant 14-08-01015), Russian Federation President's Grant MD-6325.2016.8 and St. Petersburg State University, grant 6.38.230.2015.
Theorem~\ref{thm.stab} was obtained under sole support of Russian Science Fund (RSF) grant 14-29-00142. Theorem~\ref{thm.stab-mult} was obtained under sole support of RSF grant 16-11-00063. Theorem~\ref{thm.unstab-mult} was obtained
under sole support of RSF grant grant 16-19-00057.}
}

\begin{document}

\maketitle

\begin{abstract}
Unlike many complex networks studied in the literature, social networks rarely exhibit unanimous behavior, or \emph{consensus}.
This requires a development of mathematical models that are sufficiently simple to be examined and capture, at the same time, the complex behavior of real social groups, where opinions and actions related to them may form clusters of different size. One such model, proposed by Friedkin and Johnsen, extends the idea of conventional consensus algorithm (also referred to as the iterative opinion pooling) to take into account the actors' prejudices, caused by some exogenous factors and leading to disagreement in the final opinions.

In this paper, we offer a novel multidimensional extension, describing the evolution of the agents' opinions on several topics.
Unlike the existing models, these topics are interdependent, and hence the opinions being formed on these topics are also mutually dependent.
We rigorous examine stability properties of the proposed model, in particular, convergence of the agents' opinions. Although our model assumes synchronous communication among the agents, we show that the same final opinions may be reached ``on average'' via asynchronous gossip-based protocols.
\end{abstract}

\section{Introduction}

A social network is an important and attractive case study in the theory of complex networks and multi-agent systems. Unlike many natural and man-made complex networks, whose cooperative behavior is motivated by the attainment of some global coordination among the agents, e.g. \emph{consensus}, opinions of social actors usually disagree and may form irregular factions (clusters).
We use the term ``opinion'' to broadly refer to individuals' displayed cognitive
orientations to objects (e.g., topics or issues); the term includes displayed attitudes (signed orientations) and beliefs (subjective certainties). A challenging problem is to develop a model of opinion dynamics, admitting mathematically rigorous analysis, and yet sufficiently instructive to capture the main properties of real social networks.

The backbone of many mathematical models, explaining the clustering of continuous opinions, is the idea of \emph{homophily} or \emph{biased assimilation}~\cite{Dandekar:2013}: a social actor readily adopts opinions of like-minded individuals (under the assumption that its small differences of opinion with others are not evaluated as important), accepting the more deviant opinions with discretion. This principle is prominently manifested by various \emph{bounded confidence}
models, where the agents completely ignore the opinions outside their confidence intervals \cite{Krause:2002,DeffuantWeisbuch:2000,LorenzSurvey:2007,Blondel:2009}. These models demonstrate clustering of opinions, however,
their rigorous mathematical analysis remains a non-trivial problem; it is very difficult, for instance, to predict the structure of opinion clusters for a given initial condition. Another possible explanation of opinion disagreement is \emph{antagonism} among some pairs of agents, naturally described by \emph{negative ties} \cite{Flache:2011}.
Special dynamics of this type, leading to opinion polarization, were addressed in \cite{Altafini:2012,Altafini:2013,Valcher:14,ProMatvCao:2016,ProMatvCao:2014}. It should be noticed, however, that
experimental evidence securing the postulate of ubiquitous negative interpersonal influences
(also known as \emph{boomerang effects}) seems to be
currently unavailable. Since the first definition of boomerang effects \cite{HovlandBook}, the empirical literature has concentrated
on the special conditions under which these effects might arise. This literature provides no assertion that boomerang
effects, sometimes observed in dyad systems, are non-ignorable in multi-agent networks of social influence.

It is known that even a network with positive and linear couplings may exhibit persistent disagreement and clustering, if its nodes are heterogeneous, e.g. some agents are ``informed'', that is, influenced by some external signals~\cite{XiaCao:11,AeyelsSmet:11}. One of the first models of opinion dynamics, employing such a heterogeneity, was suggested in
\cite{FriedkinJohnsen:1999,FriedkinBook,FriedkinJohnsenBook} and henceforth is referred to as the Friedkin-Johnsen (FJ) model. The FJ model promotes and extends the DeGroot iterative pooling scheme \cite{DeGroot}, taking its origins in French's ``theory of social power''~\cite{French:1956,Harary:1959}. Unlike the DeGroot scheme, where each actor updates its opinion based on its own and neighbors' opinions, in the FJ model actors can also factor their initial opinions, or \emph{prejudices}, into every iteration of opinion. In other words, some of the agents are \emph{stubborn} in the sense that they never forget their prejudices, and thus remain persistently influenced by exogenous conditions under which those prejudices were formed \cite{FriedkinJohnsen:1999,FriedkinBook}. In the recent papers \cite{FrascaTempo:2013,FrascaTempo:2015} a sufficient condition for stability of the FJ model was obtained. Furthermore, although the original FJ model is based on synchronous communication, in \cite{FrascaTempo:2013,FrascaTempo:2015} its ``lazy'' version was proposed. This version is based on asynchronous gossip influence and provides the same steady opinion \emph{on average}, no matter if one considers the probabilistic average (that is, the expectation) or time-average (the solution Ces\`aro mean).
The FJ model and its gossip modification are intimately related to the PageRank computation algorithms \cite{FriedkinJohnsen:2014,IshiiTempo:2010,IshiiTempo:2014,TempoBook,FrascaTempo:2015,FrascaIshiiTempo:2015,ProTempoCao16-1}.
In special cases, the FJ model has been given a game-theoretic~\cite{Bindel:2011} and an electric interpretation~\cite{GhaderiSrikant:2014}.
Similar dynamics arise in Leontief economic models \cite{MilaneseTempo:1988} and some protocols for multi-agent coordination~\cite{Cao:15-Opinion}. Further extensions of the FJ model are discussed in the recent papers \cite{Bullo:2013,Friedkin:2015}.

Whereas many of the aforementioned models of opinion dynamics deal with scalar opinions, we deal with influence that may modify opinions on several topics, which makes it natural to consider \emph{vector-valued} opinions \cite{Fortunato:2005,Nedic:2012,LorenzSurvey:2007,Scaglione:2013}; each opinion vector in such a model is constituted by $d>1$  topic-specific scalar opinions. A corresponding multidimensional extension has been also suggested for the FJ model \cite{FriedkinJohnsenBook,Friedkin:2015}. However, these extensions assumed that opinions' dimensions are \emph{independent},
that is, agents' attitudes to each specific topic evolve as if the other dimensions did not exist.
In contrast, if each opinion vector is constituted by an agent's opinions on several \emph{interdependent} issues, then the dynamics of the topic-specific opinions are entangled. It has long been recognized that such interdependence may exist and is important. A set of interdependent positions on multiple issues is referred to as \emph{schema} in psychology, \emph{ideology} in political science, and \emph{culture} in sociology and social anthropology; scientists more often use the terms \emph{paradigm} and \emph{doctrine}. Converse in his seminal paper \cite{Converse:1964} defined a \emph{belief system} as a ``configuration of ideas and attitudes in which elements are bound together by some form of constraints of functional interdependence''. All these closely related concepts share the common idea of an interdependent set of cognitive orientations to objects.

The main contribution of this paper is a novel multidimensional extension of the FJ model, which describes the dynamics of vector-valued opinions,
representing individuals' positions on several interdependent issues. This extension, describing the evolution of a \emph{belief system}, cannot be obtained by a replication of the scalar FJ model on each issue. For both classical and extended FJ models we obtain necessary and sufficient conditions of stability and convergence. We also develop a randomized asynchronous protocol, which provides convergence to the same steady opinion vector as the original deterministic dynamics on average. This paper significantly extends results of the  paper~\cite{Parsegov2015CDC}, which deals with a special case of the FJ model~\cite{FriedkinJohnsen:1999,FrascaTempo:2013,FrascaTempo:2015} satisfying the ``coupling condition''. This condition restricts the agent's susceptibility to neighbors' opinions to coincide with its self-weight.

The paper is organized as follows. Section~\ref{sec.prelim} introduces some concepts and notation to be used throughout the paper. In Section~\ref{sec.scalar} we introduce the scalar FJ model and related concepts; its stability and convergence properties are studied in Section~\ref{sec.stab}. A novel multidimensional model of opinion dynamics is presented in Section~\ref{sec.multi}. Section~\ref{sec.gossip} offers an asynchronous randomized model of opinion dynamics, that is equivalent to the deterministic model on average. We illustrate the results by numerical experiments in Section~\ref{sec.simul}. In Section~\ref{sec.ident} we discuss approaches to the estimation of the multi-issues dependencies from experimental data. Proofs are collected in Section~\ref{sec.proof}. Section~\ref{sec.conclu} concludes the paper.

\section{Preliminaries and Notation}\label{sec.prelim}

Given two integers $m$ and $n\ge m$, let $\overline{m:n}$ denote the set $\{m,m+1,\ldots,n\}$. Given a finite set $V$, its cardinality is denoted by $|V|$. We denote matrices with capital letters $A=(a_{ij})$, using lower case letters for vectors and scalar entries.
The symbol $\mathbbm{1}_n$ denotes the column vector of ones $(1,1,\ldots,1)^{\top}\in\r^n$, and $I_n$ is the identity matrix of size $n$.

Given a square matrix $A=(a_{ij})_{i,j=1}^n$, let $\diag A=\diag(a_{11},a_{22},\ldots,a_{nn})\in\r^{n\times n}$ stand for its main diagonal and $\rho(A)$ be its \emph{spectral radius}.
The matrix $A$ is \emph{Schur stable} if $\rho(A)<1$. The matrix $A$ is \emph{row-stochastic} if $a_{ij}\ge 0$ and $\sum_{j=1}^n a_{ij}=1\,\forall i$.
Given a pair of matrices $A\in\r^{m\times n}$, $B\in\r^{p\times q}$, their Kronecker product \cite{Laub:2005,HornJohnsonBook1991} is defined by
$$
A\otimes B=
\left[\begin{matrix}
a_{11}B & a_{12}B & \cdots & a_{1n}B\\
a_{21}B & a_{22}B & \cdots & a_{2n}B\\
\vdots & &\ddots & \vdots\\
a_{m1}B & a_{m2}B & \cdots & a_{mn}B
\end{matrix}\right]\in\r^{mp\times nq}.
$$

A (directed) \emph{graph} is a pair $\mathcal G=(\mathcal V,\mathcal E)$, where $\mathcal V$ stands for the finite set of \emph{nodes} or \emph{vertices} and
$\mathcal E\subseteq \mathcal V\times \mathcal V$ is the set of \emph{arcs} or \emph{edges}. A sequence $i=i_0\mapsto i_1\mapsto\ldots\mapsto i_r=i'$ is called a \emph{walk} from $i$ to $i'$;
the node $i'$ is \emph{reachable} from the node $i$ if at least one walk leads from $i$ to $i'$.
The graph is \emph{strongly connected} if each node is reachable from any other node. Unless otherwise stated, we assume that nodes of each graph are indexed from $1$ to $n=|\mathcal V|$,
so that $\mathcal V=\overline{1:n}$.
%We say that a matrix $A=(w_{ij})_{i,j=1}^n$ is adopted to such a graph if $a_{ij}\ne 0\Leftrightarrow (i,j)\in\mathcal E$.

\section{The FJ and DeGroot Models}\label{sec.scalar}

Consider a community of $n$ social \emph{actors} (or agents) indexed 1 through $n$, and let $x=(x_1,\ldots,x_n)^{\top}$ stand for the column vector of their scalar \emph{opinions} $x_i\in\r$. The Friedkin-Johnsen (FJ) model of opinions evolution \cite{FriedkinBook,FriedkinJohnsen:1999,FriedkinJohnsenBook} is determined by two matrices, that is
a row-stochastic matrix of \emph{interpersonal influences} $W\in \r^{n \times n}$ and a diagonal matrix of actors'
\emph{susceptibilities} to the social influence $0\le\Lambda\le I_n$ (we follow the notations from \cite{FrascaTempo:2013,FrascaTempo:2015}).
At each stage $k=0,1,\ldots$ of the influence process the agents' opinions evolve as follows
\begin{equation}\label{eq.fjmodel}
x(k+1) = \Lambda W x(k)+(I-\Lambda)u,\quad x(0)=u.
\end{equation}
The values $u_i=x_i(0)$ are referred to as the agents \emph{prejudices}.

The model~\eqref{eq.fjmodel} naturally extends DeGroot's iterative scheme of \emph{opinion pooling} \cite{DeGroot} where $\La=I_n$.
Similar to DeGroot's model, it assumes an \emph{averaging} (convex combination) mechanism of information integration. Each agent $i$ allocates weights to the displayed opinions of others under the constraint of an ongoing allocation of weight to the agent's initial opinion. The natural and intensively investigated special case of this model assumes the ``coupling condition'' $\la_{ii}=1-w_{ii}\,\forall i$ (that is, $\La=I-\diag W$). Under this assumption, the self-weight $w_{ii}$ plays a special role, considered to be a measure of \emph{stubborness} or \emph{closure} of the $i$th agent to interpersonal influence. If $w_{ii}=1$ and thus $w_{ij}=0\,\forall j\ne i$, then it is maximally stubborn and completely ignores opinions of its neighbors. Conversely, if $w_{ii}=0$ (and thus its susceptibility is maximal $\la_{ii}=1$), then the agent is completely open to interpersonal influence, attaches no weight to its own opinion (and thus forgets its initial conditions), relying fully on others' opinions. The susceptibility of the $i$th agent $\la_{ii}=1-w_{ii}$ varies between $0$ and $1$, where the extremal values correspond respectively to  maximally stubborn and open-minded agents.
From its inception, the usefulness of this special case has been empirically assessed with different measures
of opinion and alternative measurement models of the interpersonal influence matrix $W$ \cite{FriedkinBook,FriedkinJohnsen:1999,FriedkinJohnsenBook,Friedkin:2012}.

In this section, we consider dynamics of \eqref{eq.fjmodel} in the general case, where the diagonal susceptibility matrix $0\le \Lambda\le I_n$ may differ from $I-\diag W$. In the case where $w_{ii}=1$ and hence $w_{ij}=0$ as $i\ne j$, one has $x_i(1)=x_i(0)=u_i$ and then, via induction on $k$, one easily gets $x_i(k)=u_i$ for any $k=0,1,\ldots$, no matter how $\la_{ii}$ is chosen. On the other hand,
if $\la_{ii}=0$, then $x_i(k)=u_i$ independent of the weights $w_{ij}$.
Henceforth we assume, without loss of generality, that for any $i\in \overline{1:n}$ one either have $\la_{ii}=0$ and $w_{ii}=1$ (entailing that $x_i(k)\equiv u_i$) or $\la_{ii}<1$ and $w_{ii}<1$.

It is convenient to associate the matrix $W$ to the graph $\mathcal G[W]=(\mathcal V,\mathcal E[W])$. The set of nodes $\mathcal V=\overline{1:n}$ of this graph is in one-to-one correspondence with the agents
and the arcs stand for the inter-personal influences (or \emph{ties}), that is $(i,j)\in\mathcal E[W]$ if and only if $w_{ij}>0$.
A positive self-influence weight $w_{ii}>0$ corresponds to the self-loop $(i,i)$. We call $\G=\G[W]$ the \emph{interaction graph} of the social network.

\begin{figure}[h]\center
  \includegraphics[width=0.33\columnwidth]{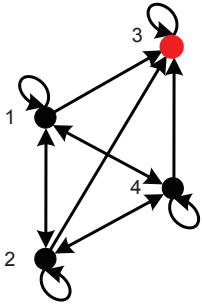}
  \caption{Interaction graph $\G[W]$, corresponding to matrix \eqref{eq.W}}
  \label{fig.graph-w}
 \end{figure}

\begin{example}\label{ex.0}
Consider a social network of $n=4$ actors, addressed in \cite{FriedkinJohnsen:1999} and having interpersonal influences as follows
\be\label{eq.W}
  W = \begin{bmatrix}
  0.220 & 0.120 & 0.360 & 0.300 \\
   0.147 & 0.215 & 0.344 & 0.294 \\
   0 & 0 & 1 & 0 \\
   0.090 & 0.178 & 0.446 & 0.286
\end{bmatrix}.
\ee
Fig.~\ref{fig.graph-w} illustrates the corresponding interaction graph.
\end{example}

In this section we are primarily interested in \emph{convergence} of the FJ model to a stationary point (if such a point exists).
\begin{definition} \textbf{(Convergence).}
The FJ model \eqref{eq.fjmodel} is \emph{convergent}, if for any vector $u\in\r^n$ the sequence $x(k)$ has a limit
\be\label{eq.x_star}
x'=\lim\limits_{k\to\infty} x(k)\Longrightarrow x'=\Lambda Wx'+(I-\Lambda)u.
\ee
\end{definition}

It should be noticed that the limit value $x'=x'(u)$ in general \emph{depends} on the initial condition $x(0)=u$. A special situation
where any solution converges to \emph{the same} equilibrium is the \emph{exponential stability} of the linear system \eqref{eq.fjmodel}, which means that $\Lambda W$ is a Schur stable matrix: $\rho(\Lambda W)<1$. A stable
FJ model is convergent, and the only stationary point is
\be\label{eq.stable-stat}
x'=\sum_{k=0}^{\infty}(\Lambda W)^k(I-\Lambda)u=(I-\Lambda W)^{-1}(I-\Lambda)u.
\ee
As will be shown, the class of convergent FJ models is in fact wider than that of stable ones. This is not surprising since, for instance, the classical DeGroot model \cite{DeGroot}
where $\Lambda=I_n$ is never stable, yet converges to a consensus value ($x'_1=\ldots=x'_n$) whenever $W$ is stochastic indecomposable aperiodic (SIA) \cite{Wolfowitz:1963}, for instance,
$W^m$ is positive for some $m>0$ (i.e. $W$ is \emph{primitive}) \cite{DeGroot}. In fact, any unstable FJ model contains a subgroup of agents whose opinions obey the DeGroot model, being independent on the remaining network.
To formulate the corresponding results, we introduce the following definition.

\begin{definition}\textbf{(Stubborness and oblivion).}
We call the $i$th agent \emph{stubborn} if $\la_{ii}<1$ and \emph{totally stubborn} if $\la_{ii}=0$.
An agent that is neither stubborn nor influenced by a stubborn agent (connected to some stubborn agent by a walk in the interaction graph $\G[W]$) is called \emph{oblivious}.
\end{definition}

\begin{figure}[h]\center
  \includegraphics[width=0.40\columnwidth]{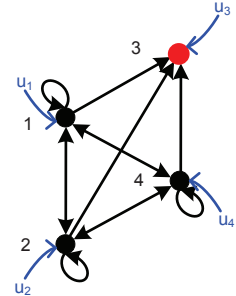}
  \caption{The structure of couplings among the agents and ``inputs'' for the FJ model with $W$ from \eqref{eq.W} and $\La=I-\diag W$}
  \label{fig.graph-wu}
 \end{figure}

\begin{example}\label{ex0cont}
Consider the FJ model \eqref{eq.fjmodel}, where $W$ is from \eqref{eq.W} and $\La=I-\diag W$.
It should be noticed that this model was reconstructed from real data, obtained in experiments with a small group of individuals, following the method proposed in \cite{FriedkinJohnsen:1999}. Figure~\ref{fig.graph-wu} illustrates the graph of the coupling matrix $\La W$ and the constant ``input'' (prejudice) $u$. In this model the agent $3$ (drawn in red) is totally stubborn, and the three agents $1$, $2$ and $4$ are stubborn. Hence, there are no oblivious agents in this model. As will be shown in the next section (Theorem~\ref{thm.stab}), the absence of oblivious agents implies stability.
\end{example}

The prejudices $u_i$ are considered to be formed by some exogenous conditions \cite{FriedkinJohnsen:1999}, and the agent's stubborness can be considered as their ongoing influence. A totally stubborn agent remains affected by those external ``cues'' and ignores the others' opinions, so its opinion is unchanged $x_i(k)\equiv u_i$. Stubborn agents, being not completely ``open-minded'', never forget their prejudices and factor them into every iteration of opinion. Non-stubborn agents are not ``anchored'' to their own prejudices yet be influenced by the others' prejudices via communication (such an influence corresponds to a walk in $G[W]$ from the agent to some stubborn individual), such individuals can be considered as ``implicitly stubborn''. Unlike them, for an oblivious agent the prejudice does not affect any stage of the opinion iteration, except for the first one.
The dynamics of oblivious agents thus depend on the ``prehistory'' of the social network only through the initial condition $x(0)=u$.

After renumbering the agents, we assume that stubborn agents and agents, influenced by them, are numbered $1$ through $n'\le n$ and oblivious agents (if they exist) have indices from $n'+1$ to $n$. For the oblivious agent $i$ we have $\la_{ii}=1$ and $w_{ij}=0\,\forall j\le n'$. Indeed, if $w_{ij}>0$ for some $j\le n'$, then the $i$th agent is
connected by a walk to some stubborn agent via agent $j$ and hence is not oblivious.
The matrices $W,\La$ and vectors $x(k)$ are therefore decomposed as follows
\be\label{eq.decomp0}
W=\begin{bmatrix}
W^{11} & W^{12}\\
0 & W^{22}
\end{bmatrix},
\La=\begin{bmatrix}
\La^{11} & 0\\
0 & I
\end{bmatrix},
x(k)=\begin{bmatrix}
x^1(k)\\
x^2(k)
\end{bmatrix},
\ee
where $x^1\in\r^{n'}$ and $W^{11}$ and $\La^{11}$ have dimensions $n'\times n'$.  If $n'=n$ then $x^2(k)$, $W^{12}$ and $W^{22}$ are absent, otherwise the oblivious agents obey the conventional
DeGroot dynamics $x^2(k+1)=W^{22}x^2(k)$, being independent on the remaining agents. If the FJ model is convergent, then the limit $W^{22}_*=\lim\limits_{k\to\infty}(W^{22})^k$ obviously
exists, in other words, the matrix $W^{22}$ is \emph{regular} in the sense of \cite[Ch.XIII, \S7]{GantmacherVol2}.
\begin{definition}\textbf{(Regularity)}
A matrix $A\in\r^{d\times d}$ is called \emph{regular} \cite{GantmacherVol2} if a limit $A_*=\lim\limits_{k\to\infty}A^k$ exists.
A regular \emph{row-stochastic matrix} $A$ is called~\emph{fully regular} \cite{GantmacherVol2} or \emph{SIA}
\cite{Wolfowitz:1963} if all rows of $A_*$ are identical, i.e. $A_*=1_dv^{\top}$, where $v\in\r^d$.
\end{definition}

In the literature, the regularity is usually defined for non-negative matrices~\cite{GantmacherVol2}, but in this paper we use this term for a general matrix. In the Appendix we examine some properties of \emph{stochastic} regular matrices, which play an important role in the convergence properties of the FJ model.

\section{Stability and convergence of the FJ model}\label{sec.stab}

The main contribution of this section is the following criterion for the convergence of the FJ model, which employs
the decomposition~\eqref{eq.decomp0}.

\begin{thm}\label{thm.stab}\textbf{(Stability and convergence)}
The matrix $\La^{11}W^{11}$ is Schur stable.
The system \eqref{eq.fjmodel} is stable if and only if there are no oblivious agents, that is, $\La W=\La^{11}W^{11}$.
The FJ model with oblivious agents is convergent if and only if $W^{22}$ is regular, i.e. the limit $W^{22}_*=\lim\limits_{k\to\infty}(W^{22})^k$ exists. In this case, the limiting
opinion $x'=\lim\limits_{k\to\infty} x(k)$ is given by
\be\label{eq.unstable-stat}
x'=
\begin{bmatrix}
(I-\Lambda^{11}W^{11})^{-1} & 0\\
0 & I
\end{bmatrix}
\begin{bmatrix}
I-\La^{11} & \La^{11}W^{12}W^{22}_*\\
0 & W^{22}_*
\end{bmatrix}u.
\ee
\end{thm}
\par
An important consequence of Theorem~\ref{thm.stab} is the stability of the FJ model, whose interaction graph is strongly connected
(or, equivalently, the matrix $W$ is \emph{irreducible} \cite{GantmacherVol2}).
\begin{cor}
If the interaction graph $\mathcal G[W]$ is strongly connected and $\La\ne I$ (i.e. at least one stubborn agent exists), then the FJ model \eqref{eq.fjmodel} is stable.
\end{cor}
\begin{IEEEproof}
The strong connectivity implies that each agent is either stubborn or connected by a walk to any of stubborn agents; hence, there are no oblivious agents.
\end{IEEEproof}

Theorem~\ref{thm.stab} also implies that the FJ model is featured by the following property. For a general system with constant input
\be\label{eq.sys0-gen}
x(k+1)=Ax(k)+Bu,
\ee
the regularity of the matrix $A$ is a necessary and sufficient condition for convergence if $Bu=0$, since $x(k)=A^kx(0)\to A_*x(0)$.
For $Bu\ne 0$, regularity is not sufficient for the existence of a limit $\lim\limits_{k\to\infty}x(k)$: a trivial counterexample is $A=I$. Iterating the equation \eqref{eq.sys0-gen} with regular $A$, one obtains
\be\label{eq.sys0-gen-converge}
x(k)=A^kx(0)+\sum_{j=1}^kA^jBu\xrightarrow[k\to\infty]{}A_*x(0)+\sum_{k=0}^{\infty}A^kBu,
\ee
where the convergence takes place if and only if the series in the right-hand side converge. The convergence criterion from
Theorem~\ref{thm.stab} implies that for the FJ model \eqref{eq.fjmodel} with $A=\La W$ and $B=I-\La$ the regularity of $A$ is
\emph{necessary and sufficient} for convergence  \cite{Friedkin:2015}; for any convergent FJ model \eqref{eq.sys0-gen-converge} holds.
\begin{cor}\label{cor.convergence}
The FJ model \eqref{eq.fjmodel} is convergent if and only if $A=\La W$ is regular. If this holds, the limit of powers $A_*$ is
\be\label{eq.a*fj}
A_*=\lim_{k\to\infty}(\La W)^k=
\begin{bmatrix}
0 & (I-\La^{11}W^{11})^{-1}\La^{11}W^{12}W^{22}_*\\
0 & W^{22}_*
\end{bmatrix},
\ee
and the series from \eqref{eq.sys0-gen-converge} (with $B=I-\La$) converge to
\be\label{eq.series}
\sum_{k=0}^{\infty}(\Lambda W)^k(I-\Lambda)u=
\begin{bmatrix}
(I-\La^{11}W^{11})^{-1}(I-\La^{11})u^1 \\
0
\end{bmatrix}.
\ee
Due to \eqref{eq.sys0-gen-converge}, the final opinion $x'$ from \eqref{eq.unstable-stat} decomposes into
\be\label{eq.unstable-stat1}
x'=A_*u+\sum_{k=0}^{\infty}(\Lambda W)^k(I-\Lambda)u.
\ee
\end{cor}
\begin{IEEEproof}
Theorem~\ref{thm.stab} implies that the matrix  $A=\La W$ is decomposed as follows
\ben
A=
\begin{bmatrix}
\La^{11}W^{11} & \La^{11}W^{12}\\
0 & W^{22}
\end{bmatrix},
\een
where the submatrix $\La^{11}W^{11}$ is Schur stable. It is obvious that $A$ is not regular unless $W^{22}$ is regular, since $A^k$ contains the right-bottom block $(W^{22})^k$. A straightforward
computation shows that
if $W^{22}$ is regular, then \eqref{eq.a*fj} and \eqref{eq.series} hold, in particular, $A$ is regular as well.
\end{IEEEproof}

Note that the first equality in \eqref{eq.stable-stat} in general \emph{fails} for unstable yet convergent FJ model, even though the series \eqref{eq.series}
converges to a stationary point of the system \eqref{eq.fjmodel} (the second equality in \eqref{eq.stable-stat} makes no sense as $I-\La W$ is not invertible).
Unlike the stable case, in the presence of oblivious agents the FJ model has multiple
stationary points for the same vector of prejudices $u$; the opinions $x(k)$ and the series \eqref{eq.series} converge to \emph{distinct} stationary points unless $W^{22}_*u^2=0$.

As shown in the Appendix, for a regular row-stochastic matrix $A$ the limit $A_*$ equals to
\be\label{eq.a*}
A_*=\lim_{k\to\infty} A^k=\lim_{\alpha\to 1}(I-\alpha A)^{-1}(1-\alpha).
\ee

Theorem~\ref{thm.stab}, combined with \eqref{eq.a*}, entails the following important approximation result.
Along with the FJ model \eqref{eq.fjmodel}, consider the following ``stubborn'' approximation
\be\label{eq.approx}
x_{\alpha}(k+1)=\alpha\La Wx_{\alpha}(k)+(I-\alpha\La)u,\quad x_{\alpha}(0)=u,
\ee
where $\alpha\in (0;1)$. Hence $\alpha\La<I$, which implies that all agents in the model \eqref{eq.approx} are stubborn, the model \eqref{eq.approx} is stable, converging to the stationary opinion
$x_{\alpha}(k)\xrightarrow[k\to\infty]{} x'_{\alpha}=(I-\alpha\La W)^{-1}(I-\alpha\La)u$.
It is obvious that $x_{\alpha}(k)\xrightarrow[\alpha\to 1]{} x(k)$ for any $k=1,2,\ldots$, a question arises if such a convergence takes place for $k=\infty$, that is, $x'_{\alpha}\to x'$ as $\alpha\to 1$. A straightforward computation, using \eqref{eq.a*} for $A=W^{22}$ and \eqref{eq.unstable-stat}, shows that this is the case whenever the original model \eqref{eq.fjmodel} is convergent. Moreover, the convergence is uniform in $u$, provided that $u$ varies in some compact set.
In this sense any \emph{convergent} FJ model can be approximated with the models, where all of the agents are stubborn ($\La<I_n$).
The proof of \eqref{eq.a*} in the Appendix allows to get explicit estimates for $\|x'_{\alpha}-x'\|$ that, however,
do not appear useful for the subsequent analysis.

\section{A multidimensional extension of the FJ model}\label{sec.multi}

In this section, we propose an extension of the FJ model, dealing with vector opinions $x_1(k),\ldots,x_n(k)\in\r^m$. The elements of each vector $x_i(k)=(x_i^1(k),\ldots,x_i^m(k))$ stand for the opinions of the $i$th agent on $m$ different issues.

\subsection{Opinions on independent issues}

In the simplest situation where agents communicate on $m$ completely unrelated issues, it is natural to assume that the particular issues $x_1^j(k),x_2^j(k),\ldots,x_n^j(k)$ satisfy the FJ model \eqref{eq.fjmodel} for any $j=1,\ldots,m$, and therefore
\begin{equation}\label{eq.fjmodel2-1}
x_i(k+1) = \lambda_{ii}\sum_{j=1}^nw_{ij}x_j(k) +(1-\la_{ii})u_i,\; u_i:=x_i(0).
\end{equation}

\begin{figure}[h]\center
  \includegraphics[width=0.9\columnwidth]{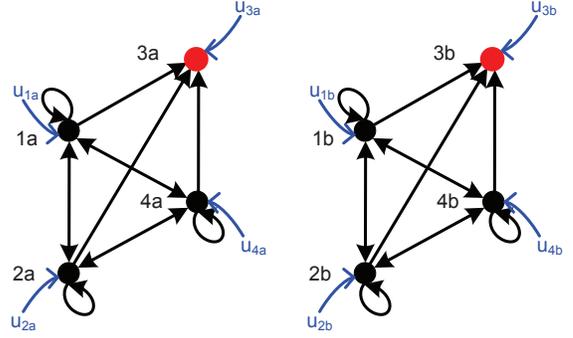}
  \caption{The structure of the two-dimensional model \eqref{eq.fjmodel2-1}}
  \label{fig.graph-independent}
 \end{figure}

\begin{example}\label{ex00}
Consider the FJ model \eqref{eq.fjmodel2-1} with $W$ from \eqref{eq.W} and $\La=I-\diag W$. Unlike Example~\ref{ex0cont}, now the opinions $x_j(k)$ are \emph{two-dimensional}, that is, $m=2$ and $x_j(k)=(x_j^1(k),x_j^2(k))^{\top}$ represent the opinions on two independent topics (a) and (b). The structure of the system, consisting of two copies of the usual FJ model \eqref{eq.fjmodel}, is illustrated by Figure~\ref{fig.graph-independent}. Since the topic-specific opinions $x_j^1(k)$, $x_j^2(k)$ evolve independently, their limits can be calculated independently, applying \eqref{eq.stable-stat} to $u^i=(x_1^i(0),x_2^i(0),x_3^i(0),x_4^i(0))^{\top}$, $i=1,2$. For instance, choosing the initial condition
\be\label{eq.initial}
x(0)=u=[\underbrace{25,25,}_{u_1=x_1(0)}\underbrace{25,15,}_{u_2=x_2(0)}\underbrace{75,-50,}_{u_3=x_3(0)}
\underbrace{85,5}_{u_4=x_4(0)}]^{\top},
\ee
the vector of steady agents' opinion is
\be\label{eq.ex-x'}
x'=[60,-19.3, 60,-21.5, 75,-50,75,-23.2]^{\top}.
\ee
\end{example}

\subsection{Interdependent issues: a belief system's dynamics}

Dealing with opinions on \emph{interdependent} topics, the opinions being formed on one topic are influenced by the opinions held on some of the other topics, so that the topic-specific opinions are entangled. Consider a group of people discussing two related topics, e.g. fish (as a part of diet) in general and salmon. Salmon is nested in fish.
A person disliking fish also dislikes salmon. If the influence process changes individuals' attitudes toward fish, say promoting fish as a healthy part of a diet, then the door is opened for influences on salmon as a part of this diet. If, on the other hand, the influence process changes individuals' attitudes against fish, say warning that fish are now contaminated by toxic chemicals, then the door is closed for influences on salmon as part of this diet.

Adjusting his/her position on one of the interdependent issues, an individual might have to adjust the positions on several related issues simultaneously in order to maintain the belief system's consistency. Contradictions and other inconsistencies between beliefs, attitudes and ideas trigger tensions and discomfort (``cognitive dissonance'') that can be resolved by a within-individual (introspective) process. This introspective process, studied in cognitive dissonance and cognitive consistency theory, is thought to be an automatic process of the human brain, enabling an individual to develop a ``coherent'' system of attitudes and beliefs~\cite{FestingerBook,GawronskiBook}.

To the best of the authors' knowledge, no model describing how networks of interpersonal influences may generate belief systems
is available in the literature. In this section, we make the first step towards filling this gap and propose a model, based on the classical FJ model, that takes issues interdependencies into account.
We modify the multidimensional FJ model \eqref{eq.fjmodel2-1} (with $x_j(k)\in\r^m$) as follows
\begin{equation}\label{eq.fjmodel2-2}
\begin{split}
x_i(k+1) &= \lambda_{ii}C\sum_{j=1}^nw_{ij}x_j(k) +(1-\la_{ii})u_i.
\end{split}
\end{equation}
The model \eqref{eq.fjmodel2-2} inherits the structure of the usual FJ dynamics, including the matrix of social influences $W$ and the matrix of agents' susceptibilities $\La$. On each stage of opinion iteration the agent $i$ calculates an ``average'' opinion, being the weighted sum $\sum_j w_{ij}x_j(k)$ of its own and its neighbors' opinions; along with the agent's prejudice $u_i$ it determines the updated opinion $x_i(k+1)$. The crucial difference with the FJ model is the presence of additional introspective transformation, adjusting and mixing the averaged topic-specific opinions. This transformation is described by a constant ``coupling matrix'' $C\in\r^{m\times m}$, henceforth called the matrix of \emph{multi-issues dependence structure} (MiDS). In the case $C=I_m$ the model \eqref{eq.fjmodel2-2} reduces to the usual FJ model \eqref{eq.fjmodel2-1}.

%Here $C$ is a row-stochastic matrix
To clarify the role of the MiDS matrix $C$, consider for the moment a network with star-shape topology where all the agents follow a totally stubborn leader, i.e. there exists $j\in\{1,2,\ldots,n\}$ such that $\la_{jj}=0$ and $w_{ij}=1=\la_{ii}$ for any $i\ne j$, so that $x_i(k+1)=Cu_j$. The opinion changes in this system are movements of the opinions of the followers toward the
initial opinions of the leader, and these movements are strictly based on the direct influences of the leader.
The entries of the MiDS matrix govern the relative contributions of the leader's issue-specific opinions to the formation of the followers' opinions. Since $x_i^p(k+1)=\sum_{q=1}^mc_{pq}u_j^q$, then $c_{pq}$ is a contribution of the $q$th issue of the leader's opinion to the $p$th issue of the follower's one.
In general, instead of a simple leader-follower network we have a group of agents, communicating on $m$ different issues in accordance with the matrix of interpersonal influences $W$.
During such communications, the $i$th agent calculates the average $\sum_jw_{ij}x_j(k)$ of its own opinion and those displayed by the neighbors. The weight $c_{pq}$ measures the effect of the $q$th issue of this averaged opinion to the $p$th issue of the updated opinion $x_i(k+1)$.

Notice that the origins and roles of matrices $W$ and $C$ in the multidimensional model \eqref{eq.fjmodel2-2} are very different. The matrix $W$ is a property of the social network, describing its topology and \emph{social influence structure}, which is henceforth assumed to be known (the measurement models for the structural matrices $\La, W$ are discussed in \cite{FriedkinBook,FriedkinJohnsen:1999,FriedkinJohnsenBook}). At the same time, $C$ expresses the interrelations between different topics of interest. It seems reasonable that the MiDS matrix should be independent of the social network itself, depending on introspective processes, forming an individual's belief system.

We proceed with examples, which show that introducing the MiDS matrix $C$ can substantially change the opinion dynamics.
These examples deal with the social network of $n=4$ actors from \cite{FriedkinJohnsen:1999}, having the influence matrix \eqref{eq.W} and the susceptibility matrix $\La=I-\diag W$. Unlike Example~\ref{ex00}, the agents discuss \emph{interdependent} topics.

\begin{example}\label{ex.1}
Let the agents discuss two topics, (a) and (b), say the attitudes towards fish (as a part of diet) in general and salmon. We start from the initial condition \eqref{eq.initial},
which means that agents $1$ and $2$ have modest positive liking for fish and salmon; the third (totally stubborn) agent has a strong liking for fish, but dislikes salmon; the agent $4$ has a strong liking for fish and a weak positive liking for salmon. Neglecting the issues interdependence ($C=I_2$), the final opinion was calculated in Example~\ref{ex00} and is given in~\eqref{eq.ex-x'}.

We now introduce a MiDS matrix, taking into account the dependencies between the topics
\begin{equation}\label{fj_correlationcc}
C_1 = \begin{bmatrix}
  0.8 & 0.2\\
   0.3 & 0.7
\end{bmatrix}.
\end{equation}
As will be shown in Theorem~\ref{thm.stab-mult}, the opinions converge to
\begin{equation}\label{eq.ex-final}
x'_{C_1}=[39.2,12,39,10.1,75,-50,56,5.3]^{\top}.
\end{equation}
Hence, introducing the MiDS matrix $C_1$ from \eqref{fj_correlationcc}, with its dominant main diagonal, imposes a substantial drag in opinions of the ``open-minded'' agents $1$ and $2$. In both cases their attitudes toward fish become more positive and those toward salmon become less positive, compared to the initial values \eqref{eq.initial}.
However, in the case of dependent issues their attitudes toward salmon do not become negative as they did in the case of independence.
As for the agent $4$, its attitude towards salmon under the MiDS matrix \eqref{fj_correlationcc} becomes even more positive, compared to the initial value \eqref{eq.initial}, whereas for $C=I_2$ this attitude becomes strongly negative.
\end{example}

\begin{figure}[h]\center
  \includegraphics[width=0.9\columnwidth]{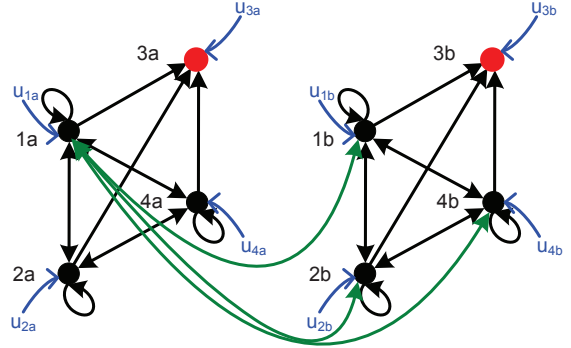}
  \caption{The structure of the two-dimensional FJ model \eqref{eq.fjmodel2-2} with $C$ from \eqref{fj_correlationcc}: emergent extra couplings between topic-specific opinions}
  \label{fig.graph-dependent}
 \end{figure}

The difference in behavior of the systems~\eqref{eq.fjmodel2-1} and~\eqref{eq.fjmodel2-2} is caused by
the presence of additional ties (couplings) between the topic-specific opinions, imposed by the MiDS matrix $C$, drawn in Fig.~\ref{fig.graph-dependent} in green. For simplicity, we show only three of these extra ties;
there are also ties between the topic-specific opinions 1b and 2a, 3a, 4a; 2a and 1b, 3b, 4b etc.

In Example~\ref{ex.1} the additional ties are positive, bringing the topic-specific opinions closer to each other. The
requirement of consistency of a belief systems may also imply the \emph{negative} couplings between different topics if the multidimensional opinion contains attitudes to a pair of contrary issues.
\begin{example}\label{ex.1+}
Consider Example~\ref{ex.1}, replacing the stochastic MiDS matrix~\eqref{fj_correlationcc} with the non-positive matrix
\be\label{fj_correlationcc+}
C_2 = \begin{bmatrix}
  0.8 & -0.2\\
   -0.3 & 0.7
\end{bmatrix}.
\ee
The topics (a) and (b), discussed by the agents, are interrelated but \emph{opposite}, e.g. the agents discuss their attitudes to
vegetarian and all-meat diets. As will be shown, the system~\eqref{eq.fjmodel2-2} remains stable.
Starting from the same initial opinion~\eqref{eq.initial}, the agents' opinions converge to the final value
\begin{equation*}\label{eq.ex-final+}
x'_{C_2}=[52.3,-30.9,52.1,-33.3,75,-50,68.4,-33.2]^{\top}.
\end{equation*}
Similar to the case of uncoupled topics~(Example~\ref{ex00}), the agents converge on positive attitude to vegetarian diets
and negative attitude to all-meat diet, following thus the prejudice of the totally stubborn individual $3$. However, the final opinions in fact substantially differ. In the case of decoupled topic-specific opinions ($C=I_2$) the agents get stronger positive attitudes to vegetarian diets than in the case of negative coupling ($C=C_2$) while their negative attitudes against all-meat diets are weaker.
\end{example}

\begin{remark}\textbf{(Restrictions on the MiDS matrix)}\label{rem.1}
In many applications the topic-specific opinions may vary only in some predefined interval. For instance,
treating them as \emph{certainties} of belief~\cite{Halpern:1991} or subjective probabilities \cite{DeGroot,Scaglione:2013},
the model of opinion evolution should imply that the topic-specific opinions belong to $[0,1]$.
Similarly, the agents \emph{attitudes}, i.e. signed orientations towards some issues~\cite{Friedkin:2015}, are
often scaled to the interval $[-1,1]$. Such a limitation on the opinions can make it natural to choose the MiDS matrix $C$ from a special class. For instance, choosing $C$ \emph{row-stochastic}, the model~\eqref{eq.fjmodel2-2} inherits important property of the FJ model~\cite{Friedkin:2015}: if $u_i^j=x_i^j(0)\in[a,b]\,\forall i,j$ then $x_i^j(k)\in[a,b]\,\forall i,j$ for any $k\ge 0$
(here $[a,b]\subset\r$ is a given interval, $i\in \overline{1:n}$ and $j\in\overline{1:m}$). This is easily derived from~\eqref{eq.fjmodel2-2} via induction on $k$: if $x_i(k)\in [a,b]^m$ for $k=0,\ldots,s$ and any $i$, then $Cx_i(k)\in [a,b]^m$ and hence $x_i(s+1)\in [a,b]^m$.

In the case of special $a$ and $b$ the assumption of row-stochasticity can be further relaxed. For instance, to keep the vector opinions $x_i(k)$ in $[0,1]^m$ whenever $u_i\in [0,1]^m$, the matrix $C$ should be chosen \emph{substochastic}: $c_{ij}\ge 0$ and $\sum_{j=1}^mc_{ij}\le 1$ for any $m$. To provide the invariance of the hypercube $[-1,1]^m$, the matrix $C$ should satisfy the condition
\be\label{eq.c-norm}
\|C\|_{\infty}=\max_i\sum_{j=1}^m|c_{ij}|\le 1.
\ee
More generally, via induction on $k$ the following invariance property can be proved: if $\mathcal D$ is a convex set and $Cx\in \mathcal D$ for any $x\in \mathcal D$, then $\mathcal D$ is an invariant set for the dynamics~\eqref{eq.fjmodel2-2}: if $u_i=x_i(0)\in\mathcal D$ then $x_i(k)\in\mathcal D$ for any $k$.
\end{remark}

 In the next subsection, the problems of stability and convergence of the model \eqref{eq.fjmodel2-2} are addressed.

\subsection{Convergence of the multidimensional FJ model}

Similar to \eqref{eq.initial}, the stack vectors of opinions $x(k)=(x_1(k)^{\top},\ldots,x_n(k)^{\top})^{\top}$ and prejudices $u=(u_1^{\top},\ldots,u_n^{\top})^{\top}=x(0)$ can be constructed. The
dynamics \eqref{eq.fjmodel2-2} now becomes
\begin{equation}\label{eq.fjmodel3}
x(k+1) = [(\Lambda W)\otimes C]x(k)+[(I_{n}-\Lambda) \otimes I_{m}],
\end{equation}
which is a convenient representation of \eqref{eq.fjmodel2-2} in the matrix form.

We begin with stability analysis of the model \eqref{eq.fjmodel3}. In the case when $C$ is row-stochastic the stability conditions
remain the same as for the initial model \eqref{eq.fjmodel}. However, the model \eqref{eq.fjmodel3} remains stable for many non-stochastic matrices, including those with exponentially unstable eigenvalues.

\begin{thm}\label{thm.stab-mult}\textbf{(Stability)}
The model~\eqref{eq.fjmodel3} is stable (i.e. $\Lambda W\otimes C$ is Schur stable) if and only if $\rho(\Lambda W)\rho(C)<1$. If this holds, then the vector of ultimate opinions is
\begin{equation}\label{eq.fjfin3}
x'_C:=\underset {k \rightarrow \infty}{\lim} x(k)=(I_{mn}-\Lambda W\otimes C)^{-1}[(I_{n}-\Lambda) \otimes I_{m}]u.
\end{equation}
If $\rho(C)=1$, the stability of~\eqref{eq.fjmodel3} is equivalent to the stability of the FJ model~\eqref{eq.fjmodel}, i.e. to the
absence of oblivious agents.
\end{thm}

\begin{remark}
Theorem~\ref{thm.stab-mult}, in particular, guarantees stability when the original FJ model~\eqref{eq.fjmodel} is stable ($\rho(\La W)<1$)
and $C$ is row-stochastic or, more generally, satisfies~\eqref{eq.c-norm}.
These conditions, however, are not necessary for stability; the system~\eqref{eq.fjmodel3} remains stable whenever $\rho(C)<\rho(\La W)^{-1}$.
\end{remark}

%\begin{IEEEproof}
%The first two claims follow from the fact that for any pair of square matrices $A,B$ the spectrum $\sigma(A\otimes B)$ coincides with elementwise product $\sigma(A)\sigma(B)$
%\cite{HornJohnsonBook1991}, and hence $\rho(A\otimes B)=\rho(A)\rho(B)$. Therefore, the matrix $\Lambda W\times C$ is Schur stable if and only if $\rho(\Lambda W)\rho(C)<1$, in which case
%the convergence \eqref{eq.fjfin3} holds. To prove the last claim, note that $\rho(C)=1$ for stochastic matrices $C$ and hence $\Lambda W\otimes C$ and $\Lambda W$ are both Schur stable or not.
%\end{IEEEproof}

In the case where some agents are oblivious, the convergence of the model ~\eqref{eq.fjmodel3} is not possible unless $C$ is regular,
that is, the limit $C_*=\lim\limits_{k\to\infty}C^k$ exists (in particular, $\rho(C)\le 1$).
As in Theorem~\ref{thm.stab}, we assume that oblivious agents are indexed $n'+1$ through $n$ and consider the decomposition~\eqref{eq.decomp0}.
\begin{thm}\label{thm.unstab-mult}\textbf{(Convergence)}
Let $n'<n$. The model~\eqref{eq.fjmodel3} is convergent if and only if $C$ is regular and either $C_*=0$ or $W^{22}$ is regular. If this holds then $x(k)\xrightarrow[k\to\infty]{}x_C'$, where
\be\label{eq.fjfin3+}
\begin{split}
x'_C&=
\begin{bmatrix}
(I-\Lambda^{11}W^{11}\otimes C)^{-1} & 0\\
0 & I
\end{bmatrix}Pu,\\
P&=\begin{bmatrix}
(I-\La^{11})\otimes I_m & (\La^{11}W^{12}W^{22}_*)\otimes CC_*\\
0 & W^{22}_*\otimes C_*
\end{bmatrix}.
\end{split}
\ee
By definition, in the case where $C_*=0$ but the limit $\lim_{k\to\infty}(W^{22})^k$ does not exist we put $W_*=0$.
\end{thm}

\begin{remark}\textbf{(Extensions)}
In the model \eqref{eq.fjmodel3} we do not assume the interdependencies between the initial topic-specific opinions;
one may also consider a more general case when $x_i(0)=Du_i$ and hence $x(0)=[I_n\otimes D]u$, where $D$ is a constant $m\times m$ matrix.
This affects neither stability nor convergence conditions, and formulas \eqref{eq.fjfin3}, \eqref{eq.fjfin3+}
for $x'_C$ remain valid, replacing $P$ in the latter equation with
\ben
P=\begin{bmatrix}
(I-\La^{11})\otimes I_m & (\La^{11}W^{12}W^{22}_*)\otimes CC_*D\\
0 & W^{22}_*\otimes C_*D
\end{bmatrix}.
\een
\end{remark}

\section{Opinion Dynamics under Gossip-Based Communication}\label{sec.gossip}

A considerable restriction of the model \eqref{eq.fjmodel3}, inherited from the original Friedkin-Johnsen model, is the \emph{simultaneous} communication between the agents. That is, on each step the actors simultaneously
communicate to all of their neighbors. This type of communication can hardly be implemented in a large-scale social network, since, as was mentioned in \cite{FriedkinJohnsen:1999},
\emph{``it is obvious that interpersonal influences do not occur in the simultaneous way and there are complex sequences of interpersonal influences in a group''}.
A more realistic opinion dynamics can be based on asynchronous \emph{gossip-based}
\cite{Boyd:06,GossipReview88} communication, assuming that
only two agents interact during each step. An asynchronous version of the FJ model \eqref{eq.fjmodel} was proposed in \cite{FrascaTempo:2013,FrascaTempo:2015}.

The idea of the model from \cite{FrascaTempo:2013,FrascaTempo:2015} is as follows. On each step an arc is randomly sampled with the uniform distribution from the interaction graph $\mathcal G[W]=(\mathcal V,\mathcal E)$. If this arc is $(i,j)$, then the $i$th agent updates its opinion in accordance with
\be\label{eq.gossip1}
x_i(k+1)=h_i\left((1-\gamma_{ij})x_i(k)+\gamma_{ij}x_j(k)\right)+(1-h_i)u_i.
\ee
Hence, the new opinion of the agent is a weighted average of his/her previous opinion, the prejudice and the neighbor's previous opinion.
The opinions of other agents remain unchanged
\be\label{eq.gossip2}
x_l(k+1)=x_l(k)\quad\forall l\ne i.
\ee

The coefficient $h_i\in [0,1]$ is a measure of the agent ``obstinacy''. If an arc $(i,i)$ is sampled, then
\be\label{eq.gossip1loop}
x_i(k+1)=h_ix_i(k)+(1-h_i)u_i.
\ee
The smaller is $h_i$, the more stubborn is the agent, for $h_i=0$ it becomes totally stubborn. Conversely, for $h_i=1$ the agent is ``open-minded'' and forgets its prejudice.
The coefficient $\gamma_{ij}\in [0,1]$ expresses how strong is the influence of the $j$th agent on the $i$th one.
Since the arc $(i,j)$ exists if and only if $w_{ij}>0$, one may assume that $\gamma_{ij}=0$ whenever $w_{ij}=0$.
% one can also put $\gamma_{ii}=0\,\forall i$ as for $i=j$ the value
%$x_i(k+1)$ is independent of $\gamma_{ii}$.

It was shown in \cite{FrascaTempo:2013,FrascaTempo:2015} that, for \emph{stable} FJ model with $\La=I-\diag W$, under proper
choice of the coefficients $h_i$ and $\gamma_{ij}$, the expectation $\E x(k)$ converges to the same steady value $x'$ as the Friedkin-Johnsen model and, moreover, the process is
\emph{ergodic} in both mean-square and almost sure sense. In other words, both probabilistic averages (expectations) and time averages
(referred to as the \emph{Ces\`aro} or \emph{Polyak} averages) of the random opinions converge to the final
opinion in the FJ model. It should be noticed that opinions themselves are \emph{not convergent} (see numerical simulations below) but oscillate around their expected values.
In this section, we extend the gossip algorithm from \cite{FrascaTempo:2013,FrascaTempo:2015} to the case where $\Lambda\ne I-\diag W$ and the opinions are multidimensional.

Let $\G[W]=(\mathcal V,\mathcal E)$ be the interaction graph of the network.
 %We say the matrix $\Gamma=(\gamma_{ij})$ is \emph{adopted} to $\G$ if $\gamma_{ij}=0$ unless $(i,j)\in\mathcal E$, i.e. $w_{ij}>0$.
Given two matrices $\Gamma^1,\Gamma^2$ such that $\gamma_{ij}^1,\gamma_{ij}^2\ge 0$ and $\gamma_{ij}^1+\gamma_{ij}^2\le 1$,
we consider the following multidimensional extension of the algorithm \eqref{eq.gossip1}, \eqref{eq.gossip2}.
On each step an arc is uniformly sampled in the set $\mathcal E$. If this arc is $(i,j)$, then agent $i$ communicates to agent $j$ and updates its opinion as follows
\be\label{eq.gossipC}
x_i(k+1)=(1-\gamma_{ij}^1-\gamma_{ij}^2)x_i(k)+\gamma_{ij}^1Cx_j(k)+\gamma_{ij}^2u_i.
\ee
Hence during each interaction the agent's opinion is averaged with its own \emph{prejudice} and modified neighbors' opinion $Cx_j(k)$.
The other opinions remain unchanged as in~\eqref{eq.gossip2}.

The following theorem shows that under the assumption of the stability of the original FJ model \eqref{eq.fjmodel3} and proper choice of $\Gamma^1,\Gamma^2$
 the model \eqref{eq.gossipC}, \eqref{eq.gossip2} inherits the asymptotic properties of the deterministic model \eqref{eq.fjmodel3}.
\begin{thm}\label{thm.gossip1}\textbf{(Ergodicity)}
Assume that $\rho(\Lambda W)<1$, i.e. there are no oblivious agents, and $C$ is row-stochastic. Let $\Gamma^1=\Lambda W$ and $\Gamma^2=(I-\Lambda)W$.
Then, the limit $x_*=\lim\limits_{k\to\infty}\E x(k)$ exists and equals to the final opinion \eqref{eq.fjfin3} of the FJ model \eqref{eq.fjmodel3}, i.e.
$x_*=x'_C$. The random process $x(k)$ is \emph{almost sure ergodic}, which means that $\bar x(k)\to x_*$ with probability $1$,
and \emph{$L^p$-ergodic} so that $\E\|\bar x(k)-x_*\|^p\xrightarrow[k\to\infty]{} 0$, where
\be\label{eq.cesaro1}
\bar x(k):=\frac{1}{k+1}\sum_{l=0}^kx(l).
\ee
Both equality $x_*=x'_C$ and ergodicity remain valid, replacing $\Gamma^2=(I-\Lambda)W$ with any matrix, such that $0\le \gamma_{ij}^2\le 1-\gamma_{ij}^1$, $\sum_{j=1}^n\gamma_{ij}^2=1-\la_{ii}$ and
$\gamma_{ij}^2=0$ as $(i,j)\not\in\mathcal E$.
\end{thm}

As a corollary, we obtain the result from \cite{FrascaTempo:2013,FrascaTempo:2015}, stating the equivalence on average between the asynchronous opinion dynamics
\eqref{eq.gossip1}, \eqref{eq.gossip2} and the scalar FJ model \eqref{eq.fjmodel}.
\begin{cor}\label{cor.gossip}
Let $d_i$ be the \emph{out-branch} degree of the $i$th node, i.e. the cardinality of the set $\{j:(i,j)\in\mathcal E\}$.
Consider the algorithm \eqref{eq.gossip1}, \eqref{eq.gossip2}, where $x_i\in\r$, $(1-h_i)d_i=1-\la_{ii}\,\forall i$, $\gamma_{ij}\in [0,1]$ and $h_i\gamma_{ij}=\la_{ii}w_{ij}$ whenever $i\ne j$.
Then, the limit $x_*=\lim\limits_{k\to\infty}\E x(k)$ exists and equals to the steady-state opinion \eqref{eq.stable-stat} of the FJ model \eqref{eq.fjmodel}:
$x_*=x'$. The random process $x(k)$ is almost sure and mean-square ergodic.
\end{cor}
\begin{IEEEproof}
The algorithm~\eqref{eq.gossip1}, \eqref{eq.gossip2} can be considered as a special case of~\eqref{eq.gossipC}, \eqref{eq.gossip2}, where $C=1$, $\gamma_{ij}^1=h_i\gamma_{ij}$ and
$\gamma_{ij}^2=1-h_i$. Since the values $\gamma_{ii}^1$ have no effect on the dynamics \eqref{eq.gossipC} with $C=1$,
one can, changing $\gamma_{ii}^1$ if necessary, assume that $\Gamma^1=\La W$. The claim now follows from Theorem~\ref{thm.gossip1} since $1-\gamma_{ij}^2=h_i\ge\gamma_{ij}^1$ and
$\sum_j\gamma_{ij}^2=(1-h_i)d_i=1-\la_{ii}$.
\end{IEEEproof}

Hence, the gossip algorithm, proposed in \cite{FrascaTempo:2013,FrascaTempo:2015} is only one element of a broad class
of protocols~\eqref{eq.gossipC} (with $C=1$), satisfying assumptions of Theorem~\ref{thm.gossip1}.

\begin{remark}\textbf{(Random opinions)}
Whereas the Ces\`aro-Polyak averages $\bar x(k)$ do converge to their average value $x_*$, the random opinions $x(k)$ themselves \emph{do not}, exhibiting non-decaying oscillations around $x_*$,
see \cite{FrascaTempo:2013} and the numerical simulations in Section~\ref{sec.simul}. As implied by \cite[Theorem~1]{FrascaTempo:2015}, $x(k)$ converges in probability to a random vector $x_{\infty}$, whose distribution is the unique invariant distribution of the dynamics \eqref{eq.gossipC},~\eqref{eq.gossip2} and is determined by the triple $(\La, W, C)$.
\end{remark}

\section{Numerical experiments}\label{sec.simul}

In this section, we give numerical tests, which illustrate the convergence of the ``synchronous'' multidimensional FJ model and its ``lazy'' gossip version.

\begin{figure}[h]\center
\begin{subfigure}[b]{0.75\linewidth}
\center
  \includegraphics[width=7cm]{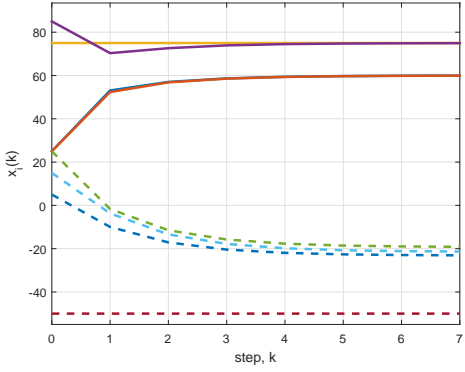}
  \caption{Example~\ref{ex00}: topics are independent}
  \label{f1}
 \end{subfigure}
 \begin{subfigure}[b]{0.75\linewidth}
 \center
  \includegraphics[width=7cm]{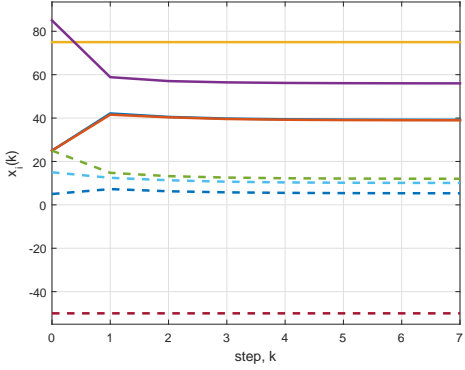}
  \caption{Example~\ref{ex.1}: topics are positively coupled}
    \label{f4}
 \end{subfigure}
  \begin{subfigure}[b]{0.75\linewidth}
  \center
  \includegraphics[width=7cm]{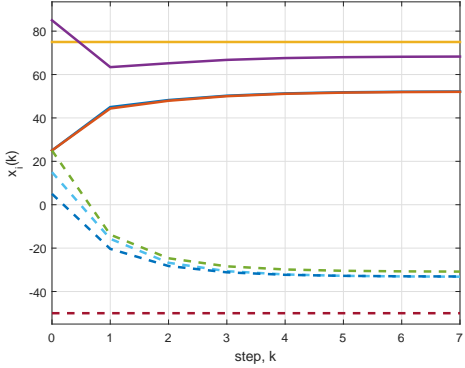}
  \caption{Example~\ref{ex.1+}: topics are negatively coupled}
    \label{f4+}
 \end{subfigure}
 \caption{Dynamics of opinions in Examples~\ref{ex00}-\ref{ex.1+}.}
 \label{f1-4-4p}
 \end{figure}

We start with the opinion dynamics of a social network with $n=4$ actors, the matrix of interpersonal influences $W$
from \eqref{eq.W} and susceptibility matrix $\La=I-\diag W$.
In Fig.~\ref{f1-4-4p} we illustrate the dynamics of opinions in Examples~\ref{ex00}-\ref{ex.1+}:
Fig.~\ref{f1} shows the case of independent issues $C=I_2$, Fig.~\ref{f4} illustrates the model~\eqref{eq.fjmodel2-2}
with the stochastic matrix $C$ from~\eqref{fj_correlationcc}, and Fig.~\ref{f4+} demonstrates the dynamics under
the MiDS matrix~\eqref{fj_correlationcc+}.
As was discussed in Examples~\ref{ex.1} and~\ref{ex.1+}, the interdependencies between the topics lead to substantial drags in the opinions of the agents 1,2 and 4, compared with the case of independent topics.

\begin{figure}[h]\center
\begin{subfigure}[b]{0.75\linewidth}\center
  \includegraphics[width=7cm]{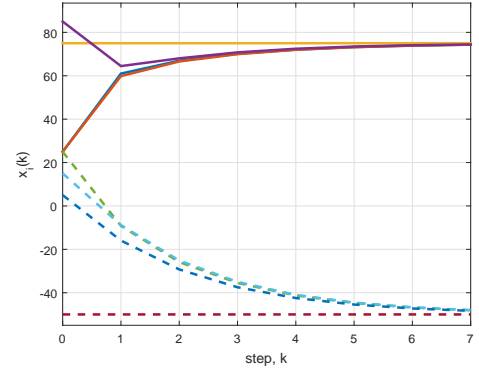}
  \caption{DeGroot's model: independent issues}
  \label{deGroot1}
 \end{subfigure}
 \begin{subfigure}[b]{0.75\linewidth}\center
  \includegraphics[width=7cm]{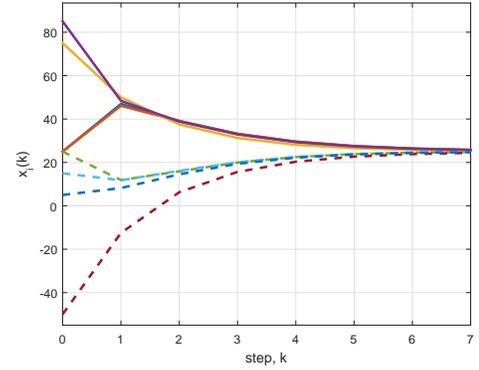}
  \caption{Extended DeGroot's model: MiDS matrix~\eqref{fj_correlationcc} }
    \label{deGroot2}
 \end{subfigure}
  \begin{subfigure}[b]{0.75\linewidth}\center
  \includegraphics[width=7cm]{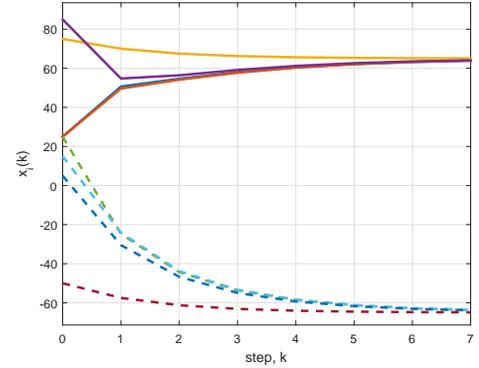}
  \caption{Extended DeGroot's model: MiDS matrix~\eqref{fj_correlationcc+}}
    \label{deGroot3}
 \end{subfigure}
 \caption{Opinion dynamics in the extended DeGroot model}
 \label{fig.deGroot}
 \end{figure}

It is useful to compare the final opinion of the models just considered with the DeGroot-like dynamics\footnote{In the DeGroot model \cite{DeGroot} the components of the opinion vectors $x_i(k)$ are
independent. This corresponds to the case where $C=I_m$. One can consider a generalized DeGroot's model as well, which is a special case of \eqref{eq.fjmodel3} with $\La=I_n$ but $C\ne I_m$. This
implies the issues interdependency, which can make all topic-specific opinions (that is, opinions on different issues) converge to the same consensus value or polarize, as shown in Fig.~\ref{fig.deGroot}.} (Fig.~\ref{fig.deGroot}) where the initial opinions and matrices $C$ are the same, however, $\La=I_n$. In the case of independent issues $C=I_2$ all the opinions
are attracted by the stubborn agent's opinion (Fig.~\ref{deGroot1})
$$
\lim\limits_{k\to\infty} x(k)=[75,-50,75,-50,75,-50,75,-50]^{\top}.
$$
In the case of positive ties between topics (Fig.~\ref{deGroot2}) we have
$$
\lim\limits_{k\to\infty} x(k)=[25,25,25,25,25,25,25,25]^{\top}.
$$
In fact, the stubborn agent $3$ constantly averages the issues of its opinions so that they reach agreement, all other issues are also attracted to this consensus value. In the case of negatively coupled topics (Fig.~\ref{deGroot3}) the topic-specific opinions \emph{polarize}
$$
\lim\limits_{k\to\infty} x(k)=[65,-65,65,-65,65,-65,65,-65]^{\top}.
$$

\begin{figure}[h]\center
  \begin{subfigure}[b]{0.75\linewidth}\center
  \includegraphics[width=7cm]{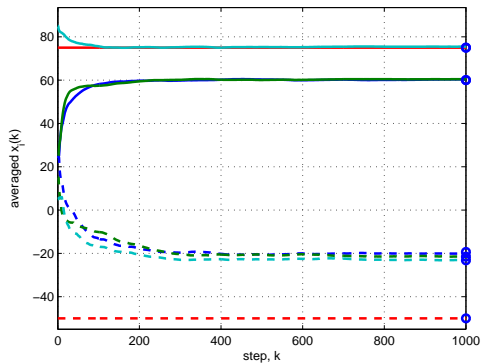}
  \caption{Convergence of the Ces\`{a}ro-Polyak averages, the MiDS matrix $C=I_2$}
  \label{f3}
 \end{subfigure}
 \begin{subfigure}[b]{0.75\linewidth}\center
 \center
  \includegraphics[width=7cm]{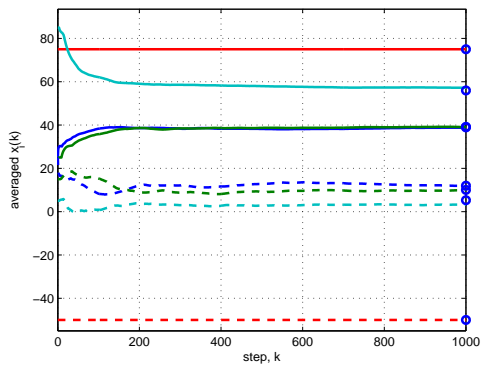}
  \caption{Convergence of the Ces\`{a}ro-Polyak averages, the MiDS matrix~\eqref{fj_correlationcc}}
    \label{f6}
 \end{subfigure}
\begin{subfigure}[b]{0.75\linewidth}\center
  \includegraphics[width=7cm]{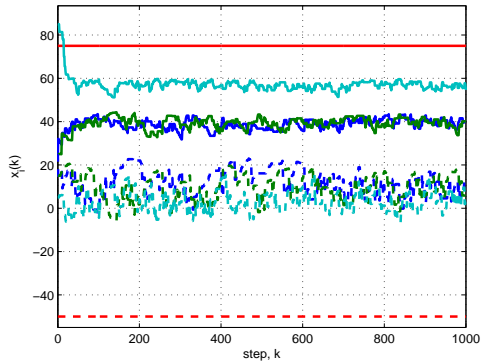}
  \caption{The dynamics of opinions with the MiDS matrix~\eqref{fj_correlationcc} show no convergence}
    \label{f5}
 \end{subfigure}
\caption{Gossip-based counterparts of Examples~\ref{ex00} and~\ref{ex.1}}\label{fig.gossip}
 \end{figure}

 The next simulation (Fig.~\ref{fig.gossip}) deals with the randomized gossip-based counterparts of the models from Examples~\eqref{ex00} and~\eqref{ex.1}. The Ces\`{a}ro-Polyak averages (Figs.~\ref{f3} and~\ref{f6}) $\bar x(k)$ of the opinions under the
gossip-based protocol from Theorem~\ref{thm.gossip1} converge to the same limits as the deterministic model~\eqref{eq.fjmodel3} (blue circles). However, the random opinions $x(k)$ themselves do not converge and exhibit oscillations (Fig.~\ref{f5}).

%\begin{figure}[h]\center
%  \includegraphics[width=7cm]{independent_issues_gossip}
%  \caption{Gossip-based dynamics with $C=I_2$, opinions}
%  \label{f2}
% \end{figure}

\begin{figure}[h]\center
  \includegraphics[width=0.75\columnwidth, height=1.0\columnwidth]{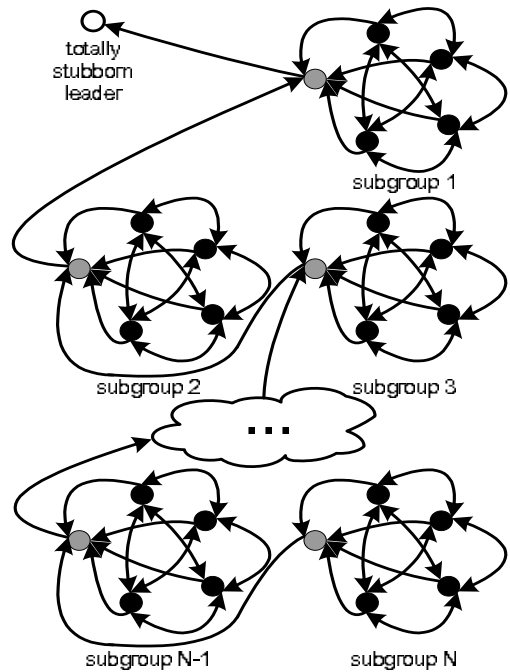}
  \caption{Hierarchical structure with $n=51$ agents}
    \label{fig.graph}
 \end{figure}

\begin{figure}
\begin{subfigure}[b]{0.75\linewidth}
  \includegraphics[width=\columnwidth]{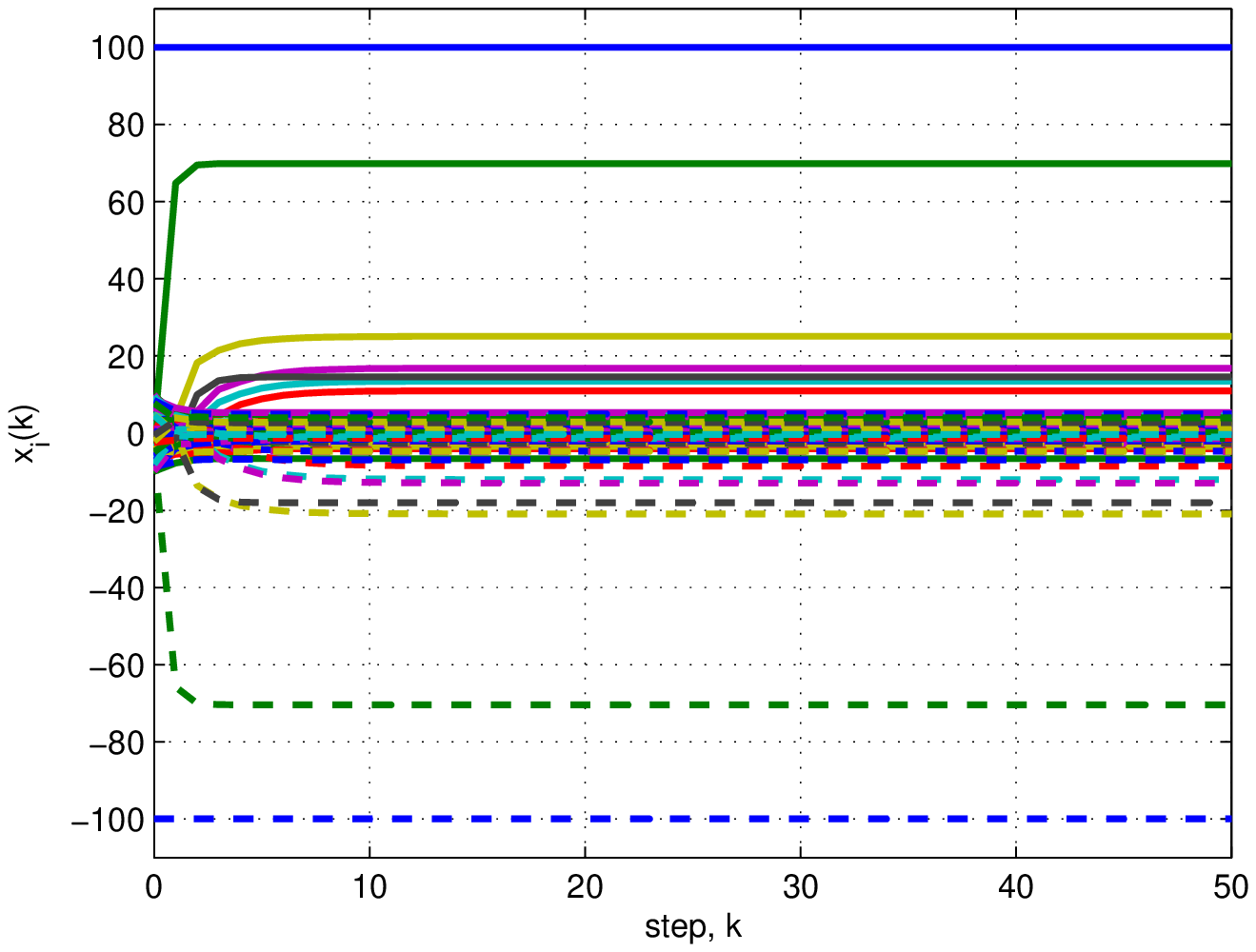}
  \caption{}
    \label{fig.large}
 \end{subfigure}
\begin{subfigure}[b]{0.75\linewidth}
  \includegraphics[width=\columnwidth]{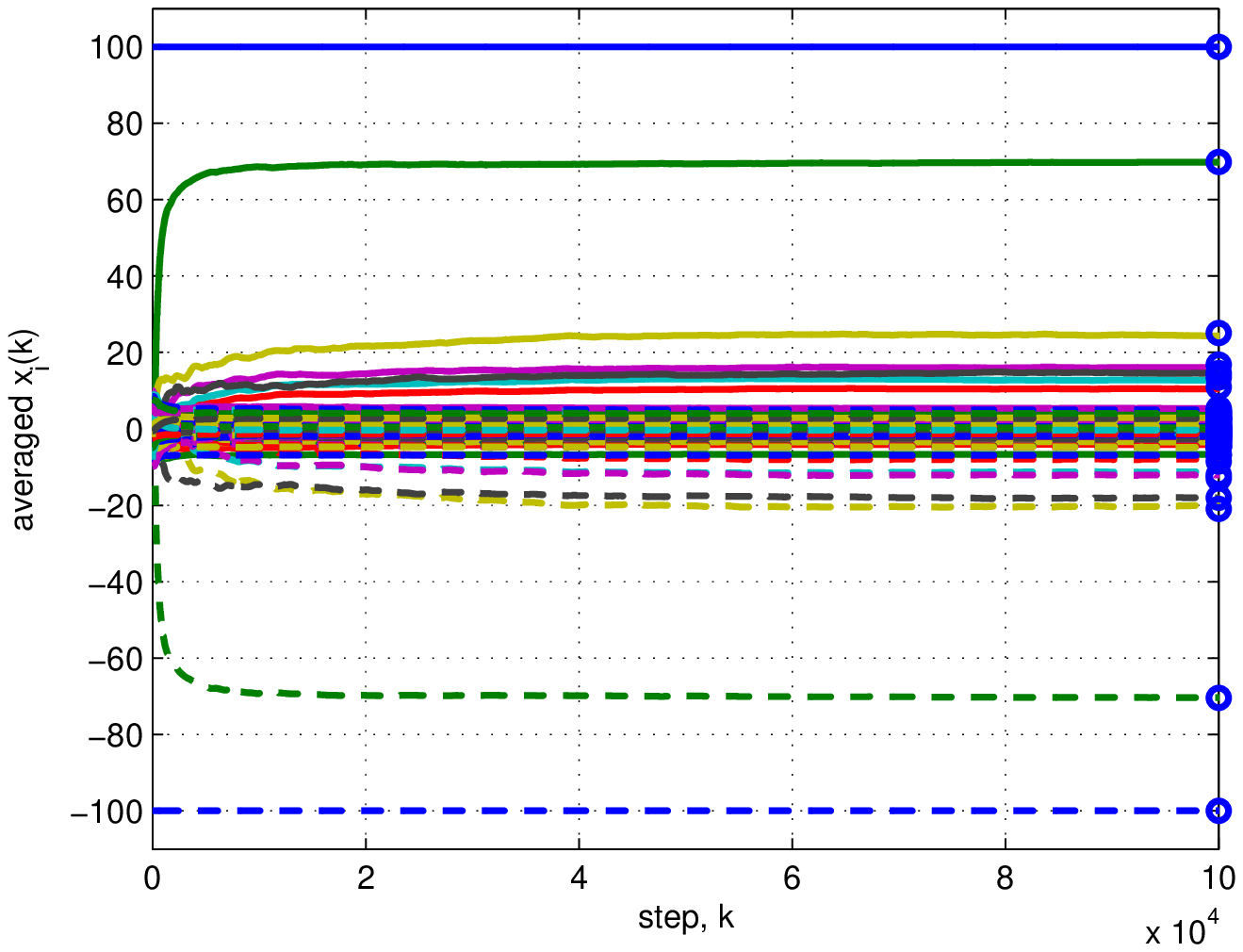}
  \caption{}
    \label{fig.large-av}
 \end{subfigure}
 \caption{Opinion dynamics of $n=51$ agents: (a) the deterministic model; (b) averaged opinions in the gossip-based model}
 \end{figure}

The last example deals with the group of $n=51$ agents, consisting of a totally stubborn ``leader'' and $N=10$ groups, each
containing $5$ agents (Fig.~\ref{fig.graph}). In each subgroup a ``local leader'' or ``representative'' exists,
who is the only subgroup member influenced from outside. The leader of the first subgroup is influenced by the totally stubborn agent,
and the leader of the $i$th subgroup ($i=2,\ldots,N$) is influenced by that of the $(i-1)$th subgroup.
All other members in each subgroup are influenced by the local leader and by each other, as shown in
Fig.~\ref{fig.graph}. Notice that each agent has a non-zero self-weight, but we intentionally do not draw self-loops around
the nodes in order to make the network structure more clear.
We simulated the dynamics of the network, assuming that the first local leader has the self-weight $0.1$ (and assigns the weight
$0.9$ to the opinion of the totally stubborn agent), and the other local leaders have self-weights $0.5$ (assigning the weight $0.5$
to the leaders of predecessing subgroups). All the weights inside the subgroups are chosen randomly in a way that $W$
is row-stochastic (we do not provide this matrix here due to space limitations).
We assume that $\La=I-\diag W$ and choose the MiDS matrix as follows
$$
C = \begin{bmatrix}
  0.9 & 0.1\\
   0.1 & 0.9
\end{bmatrix}.
$$
The initial conditions for the totally stubborn agent are $x_1(0)=[100, -100]^{\top}$, the other initial conditions are randomly
distributed in $[-10, 10]$. The dynamics of opinions in the deterministic model and averaged opinions in the gossip model
are shown respectively in Figs.~\ref{fig.large} and \ref{fig.large-av}. One can see that several clusters of opinions emerge, and
the gossip-based protocol is equivalent to the deterministic model on average, in spite of rather slow convergence.

%Notice that discussing fish and salmon, some opinion vectors are consistent and other are not, for instance, positive attitude to fish and %negative to salmon is possible, but if the fish is disliked by an individuum, he/she cannot like salmon. Suppose that initial opinions are %``feasible'' in the sense that $x_j^1(0)\ge x_j^2(0)$ (the general attitude to fish is not worse than that to salmon). A natural question %arises whether the model \eqref{eq.fjmodel3} always generates ``feasible'' opinions. Let $M=\{(x^1,x^2):x^1\ge x^2\}$ be the set of %feasible opinions. It is obvious from \eqref{eq.fjmodel2-2} that if $Cx\in M$ whenever $x\in M$ (i.e. $M$ is invariant under $C$) and %$u=x(0)\in M$, then $x_i(k)\in M$ for any $i$ and $k$. A simple check shows that $M$ is invariant under $C$ whenever %$c_{11}+c_{12}=c_{21}+c_{22}$ and
%$c_{11}-c_{21}=c_{22}-c_{12}\ge 0$, which covers both numerical tests. Generally, if the consistency of an opinion vector,
%boils down to a convex constraint $x(k)\in M$, where $M\subseteq \r^m$ is a convex set, the opinions starting at a consistent value
%$u=x(0)\in M$ will remain consistent, provided that $M$ is invariant under operator $M$.

\section{Estimation of the MiDS matrix}\label{sec.ident}

Identification of structures and dynamics in social networks, using experimental data, is an emerging area, which is actively studied by different research communities, working on statistics~\cite{Snijders:2010}, physics~\cite{WangLaiGrebogiYe:2011} and signal processing~\cite{Scaglione:16}. We assume that the structure of social influence, that is, the matrices $W$ and $\La$ are known. A procedure for their experimental identification was discussed in \cite{FriedkinJohnsen:1999,FriedkinJohnsenBook}. In this section, we focus on the estimation of the MiDS matrix $C$. As discussed in~\cite{Scaglione:16}, identification of social network is an important step from network analysis to network control and ``sensing'' (that is, observing or filtering).

To estimate $C$, an experiment can be performed where a group of individuals with given matrices $\La$ and $W$ communicates on $m$ interdependent issues. The agents are asked to record their initial opinions, constituting the vector $u=x(0)$, after which they start to communicate. The agents interact in pairs and can be separated from each other;
the matrix $W$ determines the interaction topology of the network, that is, which pairs of agent are able to interact.
Two natural types of methods, allowing to estimate $C$, can be referred to as ``finite-horizon'' and ``infinite horizon'' identification procedures.

\subsection{Finite-horizon identification procedure}

In the experiment of the first kind the agents are asked to accomplish $T\ge 1$ full rounds of conversations and record their opinions $x_j(1),\ldots,x_j(T)$ after each of these rounds, which can be grouped into stack vectors $x(1),\ldots,x(T)$.
After collection of this data, $C$ can be estimated as the matrix best fitting the equations
\eqref{eq.fjmodel3} for $0\le k<T$. Given $x(0)=u,x(1),\ldots,x(T)$, consider the optimization problem
\be\label{eq.ident1}
\begin{gathered}
\text{minimize }\quad \sum_{j=1}^{T}\|\ve_j\|_2^2\quad\text{subject to}\\
\ve_j=x(j)-\left(\La W\otimes C\,x(j-1)+(I-\La)\otimes I\,u\right).
\end{gathered}
\ee

As was discussed in Remark~\ref{rem.1}, to provide the model ``feasibility'', that is, belonging of the opinions to a given set,
it can be natural to restrict the MiDS matrix to some known set $C\in\mathcal C$, which may e.g. consist of all row-stochastic matrices
\be\label{eq.stoch}
\sum_{j=1}^{m}c_{ij}=1 \quad \forall i, \quad c_{ij} \ge 0 \quad \forall i,j,
\ee
or all matrices, satisfying~\eqref{eq.c-norm}. In both of these examples, the problem~\eqref{eq.ident1} remains a convex QP problem,
adding the constraint $C\in\mathcal C$. More generally, one may replace~\eqref{eq.ident1} with the following \emph{convex optimization} problem
\be\label{eq.ident+}
\begin{gathered}
\text{minimize }\quad f(\ve_1,\ldots,\ve_T)\quad\text{subject to}\\
\ve_j=x(j)-\left(\La W\otimes C\,x(j-1)+(I-\La)\otimes I\,u\right),\\
C\in\mathcal{C}.
\end{gathered}
\ee
Here $f$ is a convex and \emph{positive definite} function (that is, $f(\ve_1,\ldots,\ve_T)\ge 0$ and the inequality is strict unless $\ve_1=\ldots=\ve_T=0$) and $\mathcal C$ stands for a closed convex set. In the case where $f(\ve_1,\ldots,\ve_T)=\sum_{j=1}^T|\ve_j|$ or $f(\ve_1,\ldots,\ve_T)=\max_j|\ve_j|$ the problem~\eqref{eq.ident+} becomes a standard linear programming problem.
\begin{remark}\label{rem.exist}
In the case where $\mathcal C$ is a (non-empty) \emph{compact} set, the minimum in~\eqref{eq.ident+} exists whenever $f$ is continuous. Moreover, if $f$ stands for a \emph{strictly convex} norm~\cite[Section 8.1]{BoydBook} on $\r^{T}$ then the minimum exists even for unbounded closed convex sets $\mathcal C$ since the set of all possible vectors $\{(\ve_1,\ldots,\ve_T)\}$ is closed and convex, and the optimization problem~\eqref{eq.ident+} boils down to the projection on this set. For instance, the minimum in the unconstrained optimization problem~\eqref{eq.ident1} is always achieved.
\end{remark}

\subsection{Infinite-horizon identification procedure}

The experiment of the second kind is applicable only to stable models, which inevitably implies a restriction on the MiDS matrix $C$.
As in the previous section, we suppose that $C\in\mathcal C$, where $\mathcal C$ is convex closed set of matrices. We suppose also
that all elements of $\mathcal C$ satisfy Theorem~\ref{thm.stab-mult}, that is, $\rho(C)<\rho(\La W)^{-1}$.
As discussed in Remark~2, the set $\mathcal C$ may consist e.g. of all stochastic matrices~\eqref{eq.stoch} or matrices, satisfying~\eqref{eq.c-norm}.

The agents are not required to trace the history of their opinions, and their interactions are not limited to any prescribed number of rounds. Instead, similar to the experiments from \cite{FriedkinJohnsen:1999}, the agents interact until their opinions stabilize (\emph{``agents communicate until consensus or deadlock is reached''} \cite{FriedkinJohnsen:1999}). In this sense, one may assume that the agents know the final opinion $x'$. The matrix $C$ should satisfy the equation
\be\label{eq.aux1}
x'=\La W\otimes C\,x'+(I-\La)\otimes I\,u,
\ee
obtained as a limit of \eqref{eq.fjmodel3} as $k\to\infty$. To find a matrix $C$ best fitting~\eqref{eq.aux1} we
solve an optimization problem similar to~\eqref{eq.ident1}
\begin{equation}\label{eq.ident2}
\begin{gathered}
\text{minimize }\quad \|\ve\|_2^2\quad\text{subject to}\\
\ve=x'-\left(\Lambda W\otimes C\,x'+(I_n-\Lambda)\otimes I_m\,u\right),\\
C\in\mathcal C.
\end{gathered}
\end{equation}
In the case where $\mathcal C$ stands for the polyhedron of matrices (e.g. we are confined to row-stochastic MiDS matrices $C$),
the problem \eqref{eq.ident2} boils down to a convex QP problem; replacing $\|\cdot\|_2^2$ with $\|\cdot\|_1$ or $\|\cdot\|_{\infty}$ norm, one gets a LP problem. The solution in the optimization problem~\eqref{eq.ident2} exists for any closed and convex set $\mathcal C$ for the reasons explained in Remark~\ref{rem.exist}.

Both types of experiments thus lead to convex optimization problems. The advantage of the ``finite-horizon'' experiment is its independence of the system convergence. Also, allocating some fixed time for each dyadic interaction (and hence for the round of interactions), the data collection can be accomplished in known time (linearly depending on $T$). In many applications, one is primarily interested in the opinion dynamics on a finite interval. This approach, however, requires to store the whole trajectory of the system, collecting thus a large amount of data (growing as $nT$) and leads to a larger convex optimization problem. The loss of data from one of the agents in general requires to restart the experiment. On the other hand, the ``infinite-horizon'' experiment is applicable only to stable models, and one cannot predict how fast the convergence will be. This experiment does not require agents to trace their history and thus reduces the size of the optimization problem.

\subsection{Transformation of the equality constraints}

Both optimization problems \eqref{eq.ident1} and \eqref{eq.ident2} are featured by non-standard linear constraints, involving Kronecker products. To avoid Kronecker operations, we transform the constraints into a standard form $Ax=b$, where $A$ is a matrix and $x,b$ are vectors. To this end, we perform a vectorization operation. Given a matrix $M$, its \emph{vectorization} $\vec M$ is a column vector obtained by stacking the columns of $M$, one on top of one another \cite{Laub:2005}, e.g.
$\vec\left(\begin{smallmatrix}
1 & 0\\2 & 1
\end{smallmatrix}\right)=[1, 2, 0, 1]^{\top}$.
\begin{lemma}\cite{Laub:2005}
For any three matrices $\mathcal{A, B, C}$ such that the product $\mathcal{ABC}$ is defined, one has
\begin{equation}\label{laubtheorem}
\vec{\mathcal{ABC}} = (\mathcal{C}^{\top} \otimes \mathcal{A}) \vec \mathcal{B}.
\end{equation}
In particular, for $\mathcal{A} \in \mathbb{R}^{m \times l}$ and $\mathcal{B} \in \mathbb{R}^{l \times n}$ one obtains
\begin{equation}\label{eq.cor}
\vec \mathcal{AB} = (I_{n} \otimes \mathcal{A}) \vec \mathcal{B} = (\mathcal{B}^{\top} \otimes I_{m}) \vec \mathcal{A}.
\end{equation}
\end{lemma}
\vskip 3mm
The constraints in \eqref{eq.ident1}, \eqref{eq.ident2} can be simplified. Consider first the constraint in \eqref{eq.ident2}.
Let $x'_i$ be the final opinion of the $i$th agent and $X'=[x_1',\ldots,x_n']$ be the matrix constituted by them, hence
$x'=\vec X'$. Applying \eqref{eq.cor} for $\mathcal A=C$ and $\mathcal B=X'$ entails that $[I_n\otimes C]x'=[(X'^{\top}\otimes I_m]\vec C$, thus
$[\Lambda W\otimes C]x'=[\Lambda W\otimes I_m][I_n\otimes C]x'=[\Lambda W(X')^{\top}\otimes I_m]\vec C$. Denoting $c=\vec C$, the constraint in \eqref{eq.ident2} shapes into
\begin{equation}\label{eq.epsilon1}
\ve+[\Lambda WX'^{\top}\otimes I_m]c=x'-[(I_n-\Lambda)\otimes I_m]u,
\end{equation}
where the matrix $\Lambda W\hat X^{\top}\otimes I_m$ and the right-hand side are known.
The constraints in \eqref{eq.ident1} can be rewritten as
\begin{equation}\label{eq.epsilon1}
\ve_j+[\Lambda WX(j-1)^{\top}\otimes I_m]c=x(j)-[(I_n-\Lambda)\otimes I_m]u.
\end{equation}
Here $X(j)$ is the matrix $[x_1(j),\ldots,x_n(j)]$ and hence $x(j)=\vec X(j)$.

\subsection{Numerical examples}
To illustrate the identification procedures, we consider two numerical examples.
\begin{example}\label{ex.inf}
Consider a social network with the matrix $W$ from \eqref{eq.W}, $\La=I-\diag W$ and the prejudice vector \eqref{eq.initial}. Unlike Example~\ref{ex.1}, $C$ is unknown and is to be found from the ``infinite-horizon'' experiment. Suppose that agents were asked to compute the final opinion, obtaining
$$
x' = [35, 11, 35, 10, 75, -50, 53, 5]^{\top}.
$$
Solving the problem \eqref{eq.ident2},
one gets the minimal residual $\|\varepsilon\|_2 = 0.9322$, which gives the MiDS matrix
$$
C = \begin{bmatrix}
  0.7562 & 0.2438\\
  0.3032 & 0.6968
\end{bmatrix}.
$$
In accordance with \eqref{eq.fjfin3}, this matrix $C$ corresponds to the steady opinion
$$
x'_C= [35.316,   11.443,   35.092,    9.483,   75,  -50,   52.386,    4.915]^{\top}.
$$
\end{example}

\begin{example}\label{ex.fin}
For $\La$, $W$ and $u$ from the previous example, agents were asked to conduct $T=3$ full rounds of conversation (``finite horizon'' experiment), obtaining the opinions
\begin{align*}
x(1)&=[42.80,   14.05,   43.59,   12.51,   75,  -50,   61.49,    7.18]^{\top}\\
x(2)&=[41.31,   13.37,   41.45,   11.43,   75,  -50,   55.48,    6.45]^{\top}\\
x(3)&=[41.74,   12.30,   40.41,   10.84,   75,  -50,   58.99,    6.02]^{\top}.
\end{align*}
We are interested in finding a row-stochastic matrix $C$ best fitting the data. Solving the corresponding QP problem \eqref{eq.ident1}
with additional constraints~\eqref{eq.stoch}, the MiDS matrix is
$$
C = \begin{bmatrix}
  0.8181 & 0.1819\\
   0.2983 & 0.7017
\end{bmatrix}.
$$
The model~\eqref{eq.fjmodel2-2} with this matrix $C$ gives the opinions
\begin{align*}
\tilde x(1)&=[43.12,   14.66,   42.54,   12.37,   75,  -50,   59.90,    7.17]^{\top}\\
\tilde x(2)&=[41.93,   13.26,   41.73,   11.35,   75,  -50,   58.37,    6.26]^{\top}\\
\tilde x(3)&=[41.30,   12.69,   41.12,   10.79,   75,  -50,   57.90,    5.83]^{\top}.
\end{align*}
\end{example}

\begin{remark}
Examples~\ref{ex.inf} and \ref{ex.fin}, demonstrating the estimation procedures, are constructed as follows.
We get the model \eqref{eq.fjmodel3} with $W$ from \eqref{eq.W}, $\La=I-\diag W$ and $C$ from \eqref{fj_correlationcc} and slightly change its final value \eqref{eq.ex-final} (Example~\ref{ex.inf}) and trajectory (Example~\ref{ex.fin}). Due to these perturbations, the estimated MiDS matrix does not exactly coincide with \eqref{fj_correlationcc} yet is close to it.
\end{remark}

\section{Proofs}\label{sec.proof}

We start with the proof of Theorem~\ref{thm.stab}, which requires some additional techniques.
\begin{definition}\textbf{(Substochasticity)}
A non-negative matrix $A=(a_{ij})$ is \emph{row-substochastic}, if $\sum_ja_{ij}\le 1\,\forall i$. Given such a matrix sized $n\times n$,
we call a subset of indices $J\subseteq \overline{1:n}$ \emph{stochastic} if
the corresponding submatrix $(a_{ij})_{i,j\in J}$ is row-stochastic, i.e. $\sum\limits_{j\in J}a_{ij}=1\,\forall i\in J$.
\end{definition}

The Gerschgorin Disk Theorem \cite{GantmacherVol2} implies that any substochastic matrix $A$ has $\rho(A)\le 1$. Our aim is to identify the class of substochastic matrices with $\rho(A)=1$. As it will be shown, such matrices are either row-stochastic or
contain a row-stochastic submatrix, i.e. has a non-empty stochastic subset of indices.

\begin{lemma}\label{lem.tech}
Any square substochastic matrix $A$ with $\rho(A)=1$ admits a non-empty stochastic subset of indices. The union of two stochastic subsets is stochastic again, so that the \emph{maximal} stochastic subset
$J_*$ exists. Making a permutation of indices such that $J_*=\overline{(n'+1):n}$, where $0\le n'<n$, the matrix $A$ is decomposed into upper triangular form
\be\label{eq.decompose}
A=\begin{pmatrix}
A^{11} & A^{12}\\
0 & A^{22}
\end{pmatrix},
\ee
where $A^{11}$ is a Schur stable $n'\times n'$-matrix ($\rho(A^{11})<1$) and $A^{22}$ is row-stochastic.
\end{lemma}
\begin{IEEEproof}
Thanks to the Perron-Frobenius Theorem \cite{GantmacherVol2,HornJohnsonBook1991}, $\rho(A)=1$ is an eigenvalue of $A$, corresponding to a non-negative eigenvector $v\in\r^n$ (here $n$ stands for the dimension of $A$).
Without loss of generality, assume that $\max_i v_i=1$. Then we either have $v_i=1_n$ and hence $A$ is row-stochastic (so the claim is obvious),
or there exists a non-empty set $J\subsetneq \overline{1:n}$ of such indices $i$ that $v_i=1$. We are going to show that $J$ is stochastic. Since $v_i=1$ for $i\in J^c=\overline{1:n}\setminus J$, one has
$$
1=\sum_{j\in J^c}a_{ij}v_j+\sum_{j\in J}a_{ij}\le 1\forall i\in J.
$$
Since $v_j<1$ as $j\in J^c$, the equality holds only if $a_{ij}=0\,\forall i\in J,j\not\in J$ and $\sum_{j\in J}a_{ij}=1$, i.e. $J$ is a stochastic set.
This proves the first claim of Lemma~\ref{lem.tech}.

Given a stochastic subset $J$, it is obvious that $a_{ij}=0$ when $i\in J$ and $j\not\in J$, since otherwise one would have $\sum\limits_{j\in \overline{1:n}}a_{ij}>1$. This implies that given two
stochastic subsets $J_1,J_2$ and choosing $i\in J_1$, one has
$
\sum\limits_{j\in J_1\cup J_2}a_{ij}=\sum\limits_{j\in J_1}a_{ij}+\sum\limits_{j\in J_2\cap J_1^c}a_{ij}=1.
$
The same holds for $i\in J_2$, which proves stochasticity of the set $J_1\cup J_2$. This proves the second claim of Lemma~\ref{lem.tech} and the existence of the maximal stochastic subset
$J_*$, which, after a permutation of indices, becomes as follows $J_*=\overline{(n'+1):n}$. Recalling that $a_{ij}=0\,\forall i\in J_*,j\in J_*^c$, one shows that the matrix is decomposed as
\eqref{eq.decompose}, where $A^{22}$ is row-stochastic. It remains to show that $\rho(A^{11})<1$. Assume, on the contrary, that $\rho(A^{11})=1$. Applying the first claim of
Lemma~\ref{lem.tech} to $A^{11}$, one proves the existence of another stochastic subset $J'\subseteq \overline{1:n'}$, which contradicts the maximality of $J_*$. This contradiction shows that
$A^{11}$ is Schur stable.
\end{IEEEproof}

Returning to the FJ model \eqref{eq.fjmodel}, it is easily shown now that the maximal stochastic subset of indices of the matrix $\La W$ consists of indices of
\emph{oblivious} agents.
\begin{lemma}\label{lem.obliv}
Given a FJ model \eqref{eq.fjmodel} with the matrix $\La$ diagonal (where $0\le\la_{ii}\le 1$) and the matrix $W$ row-stochastic, the maximal stochastic set of indices
$J_*$ for the matrix $\La W$ is constituted by the indices of oblivious agents. In other words, $j\in J_*$ if and only if the $j$th agent is oblivious.
\end{lemma}
\begin{IEEEproof}
Notice, first, that the set $J_*$ consists of oblivious agents. Indeed, $1=\sum_{j\in J_*}\la_{ii}w_{ij}\le\la_{ii}\le 1$ for any $i\in J_*$, and hence none of agents from $J_*$ is stubborn.
Since $a_{ij}=0\,\forall i\in J_*,j\in J_*^c$ (see the proof of Lemma~\ref{lem.tech}), the agents from $J_*$ are also unaffected by stubborn agents, being thus oblivious.
Consider the set $J$ of \emph{all} oblivious agents, which, as has been just proved, comprises $J_*$: $J\supseteq J_*$.
By definition, $\la_{jj}=1\,\forall j\in J$. Furthermore, no walk in the graph from $J$ to $J^c$ exists,
and hence $w_{ij}=0$ as $i\in J,j\not\in J^c$, so that $\sum_{j\in J}w_{ij}=1\,\forall i\in J$. Therefore, indices of oblivious agents constitute a stochastic set $J$, and hence $J\subseteq J_*$.
Therefore $J=J_*$, which finishes the proof.
\end{IEEEproof}

We are now ready to prove Theorem~\ref{thm.stab}.
\begin{IEEEproof}[Proof of Theorem ~\ref{thm.stab}]
Applying Lemma~\ref{lem.tech} to the matrix $A=\La W$, we prove that agents can be re-indexed in a way that $A$ is decomposed as \eqref{eq.decompose}, where
$A^{11}=\La^{11}W^{11}$ is Schur stable and $A^{22}$ is row-stochastic (if $A$ is Schur stable, then $A=A^{11}$ and $A^{22}$ and $A^{12}$ are absent).
Lemma~\ref{lem.obliv} shows that indices $\overline{1:n'}$ correspond to stubborn agents and agents they influence, whereas indices $\overline{(n'+1):n}$ denumerate oblivious agents that are, in particular, not stubborn and hence
$\la_{jj}=1$ as $j>n'$ so that $A^{22}=W^{22}$. This proves the first claim of Theorem~\ref{thm.stab}, concerning the Schur stability of $\La^{11}W^{11}$.

By noticing that $x^2(k)=(W^{22})^kx^2(0)$, one shows that convergence of the FJ model is possible only when $W^{22}$ is regular, i.e. $(W^{22})^k\to W^{22}_*$ and hence $x^2(k)\to W^{22}_*u^2$.
If this holds, one immediately obtains \eqref{eq.unstable-stat} since
$$
x^1(k+1)=\La^{11}W^{11}x^1(k)+\La^{11}W^{12}x^2(k)+(I-\La^{11})u^1
$$
and $\La^{11}W^{11}$ is Schur stable.
\end{IEEEproof}

The proof of Theorem~\ref{thm.stab-mult} follows from the well-known property of the Kronecker product.
\begin{lemma}\label{lem.spect}\cite[Theorem 13.12]{Laub:2005}
The spectrum of the matrix $A\otimes B$ consists of all products $\la_i\mu_j$, where $\la_1,\ldots,\la_n$ are eigenvalues of $A$ and $\mu_1,\ldots,\mu_m$ are those of $B$.
\end{lemma}

\begin{IEEEproof}[Proof of Theorem~\ref{thm.stab-mult}]
Lemma~\ref{lem.spect} entails that $\rho(\La W\otimes C)=\rho(\La W)\rho(C)$, hence the system \eqref{eq.fjmodel3} is stable if and only if $\rho(\La W)\rho(C)<1$.
In particular, if $\rho(C)=1$ then the system \eqref{eq.fjmodel3} is stable if and only if $\rho(\La W)<1$.
\end{IEEEproof}

The proof of Theorem~\ref{thm.unstab-mult} is similar to that of Theorem~\ref{thm.stab}. After renumbering the agents, one can assume that oblivious agents are indexed $n'+1$ through $n$ and consider
the corresponding submatrices $W^{11},W^{12},W^{22},\La^{11}$, used in Theorem~\ref{thm.stab}. Then the matrix $\La W\otimes C$ can also be decomposed
\be\label{eq.decomposeC}
\La W\otimes C=\begin{pmatrix}
\La^{11}W^{11}\otimes C & \La^{11}W^{12}\otimes C\\
 0 & W^{22}\otimes C
\end{pmatrix},
\ee
where the matrices $\La^{11}W^{11}\otimes C$ has dimensions $mn'\times mn'$ and $m(n-n')\times m(n-n')$ respectively. We consider the corresponding subdivision of the vectors
$x(k)=[x^1(k)^{\top},x^2(k)^{\top}]^{\top}$ and $\hat u=[(\hat u^1)^{\top},(\hat u^2)^{\top}]^{\top}$.

\begin{IEEEproof}[Proof of Theorem~\ref{thm.unstab-mult}]
Since the opinion dynamics of oblivious agents is given by $x^2(k+1)=W^{22}\otimes Cx^2(k)$,
the convergence implies regularity of the matrix $W^{22}\otimes C$.
The regularity of $W^{22}\otimes C$ entails that $C$ is regular. Indeed,
consider a left eigenvector $z$ of $W^{22}$ at $1$ (that is, $zW=z$) and denote $v=z\otimes I_m$.
Since $v^T(W^{22}\otimes C)^k=z\otimes C^k$ has a limit as $k\to\infty$ and $z\ne 0$, $C$ is regular and, in particular, $\rho(C)\le 1$.
Obviously, the limit $C_*=\lim_{k\to\infty} C^k$ is zero if and only if $C$ is Schur stable, i.e. $\rho(C)<1$.
If $\rho(C)=1$ then $\la=1$ the only possible eigenvalue on the unit circle $\{\la:|\la|=1\}$.
  Consider a right eigenvector $z$ of $C$ (possibly, complex), corresponding to this eigenvalue. Denoting $v=I_n\otimes z$, the matrix $(W^{22}\otimes C)^kv=(W^{22})^k\otimes v$ has a limit as $k\to\infty$, and thus $W^{22}$ is regular.
The necessity part is proved. To prove sufficiency, notice that $x^2(k)\to W^{22}_*\otimes C_*u^2$ as $k\to\infty$ (where we put $W_*=0$ when $C_*=0$, and thus~\eqref{eq.fjfin3+} follows from the equation
\ben
\begin{split}
x^1(k+1)=[\La^{11}W^{11}\otimes C]x^1(k)+[\La^{11}W^{12}\otimes C]x^2(k)+\\+[I-\La^{11}]\otimes I_m\,u^1,
\end{split}
\een
where $\La^{11}W^{11}\otimes C$ is Schur stable.
\end{IEEEproof}

To proceed with the proof of Theorem~\ref{thm.gossip1}, we need some extra notation.
As for the scalar opinion case in \cite{FrascaTempo:2013,FrascaTempo:2015} the gossip-based protocol \eqref{eq.gossipC}, \eqref{eq.gossip2} shapes into
\be\label{eq.gossip-ab}
x(k+1)=P(k)x(k)+B(k)u,
\ee
where $P(k)$, $B(k)$ are independent identically distributed (i.i.d.) random matrices. If arc $(i,j)$ is sampled, then $P(k)=A^{(i,j)}$ and $B(k)=B^{(i,j)}$, where
by definition
\ben
\begin{split}
P^{(i,j)}&=\left(I_{mn}-(\gamma_{ij}^1+\gamma_{ij}^2)e_ie_i^{\top}\otimes I_m+\gamma_{ij}^1e_ie_j^{\top}\otimes C\right),\\
B^{(i,j)}&=\gamma_{ij}^2e_ie_i^{\top}\otimes I_m.
\end{split}
\een
Denoting $\alpha:=|\mathcal E|^{-1}\in (0,1]$ and noticing that $\E P(k)=\alpha\sum_{(i,j)\in\mathcal E}P^{(i,j)}$ and $\E B(k)=\alpha\sum_{(i,j)\in\mathcal E}B^{(i,j)}$,
the following equalities are easily obtained
\be\label{eq.ab-fj1}
\begin{split}
\E P(k)&=I_{mn}-\alpha\left[I_{mn}-\Lambda W\otimes C\right]\\
\E B(k)&=\alpha(I_n-\Lambda)\otimes I_m.
\end{split}
\ee
%that yield, finally, in
%\be\label{eq.expect1}
%\E A(k)=(1-\alpha)I+\alpha \Lambda W\otimes C, \E B(k)u=\alpha(I-\Lambda)\otimes I_m u.
%\ee

\begin{IEEEproof}[Proof of Theorem~\ref{thm.gossip1}]
As implied by equations \eqref{eq.gossip-ab} and \eqref{eq.ab-fj1}, the opinion dynamics obeys the equation
\be\label{eq.p}
x(k+1)=P(k)x(k)+v(k),
\ee
where the matrices $P(k)$ and vectors $v(k)$ are i.i.d. and their finite first moments are given by the following
\ben
\E P(k)=(1-\alpha)I+\alpha \Lambda W\otimes C,\,\E v(k)=\alpha (I_n-\Lambda)\otimes I_m\,u,
\een
where $\alpha\in (0;1]$. Since $P(k)$ are non-negative,
Theorem~1 from \cite{FrascaTempo:2015} is applicable to \eqref{eq.p}, entailing that the process $x(k)$ is almost sure ergodic and
$\E x(k)\to x_*$ as $k\to\infty$, where
$$
x_*=[I-\La W\otimes C]^{-1}[(I_n-\La)\otimes I_m]u=x'_C.
$$
To prove the $L^p$-ergodicity, notice that $x(k)$ and $\bar x(k)$ remain bounded due to Remark~\ref{rem.1}, and
hence $\E\|\bar x(k)-x_*\|^p\to 0$ thanks to the Dominated Convergence Theorem \cite{ShiriaevBook}.
\end{IEEEproof}

\begin{remark}\textbf{(Convergence rate)}
For the case of $p=2$ (mean-square ergodicity) there is an elegant estimate for the convergence rate \cite{FrascaTempo:2013,FrascaIshiiTempo:2015}:
$\E\|\bar x(k)-x_*\|^2\le \chi/(k+1)$, where $\chi$ depends on the spectral radius
$\rho(\La W)$ and the vector of prejudices $u$. An analogous estimate can be proved for the multidimensional gossip algorithm \eqref{eq.gossipC}, \eqref{eq.gossip2}.
\end{remark}

\begin{remark}\textbf{(Relaxation of the stochasticity condition)}
As can be seen from the proof, Theorem~\ref{thm.gossip1} retains its validity for substochastic matrices, since they are non-negative and provide boundedness of the solutions. Furthermore, the proof of \emph{almost sure} ergodicity does not rely on the solutions' boundedness and hence is preserved whenever $C$ is non-negative and $\rho(C)\rho(\La W)<1$. A closer examination of the proof of \cite[Theorem~1]{FrascaTempo:2015} shows that it can be extended to the case where $P(k)$ with negative entries.
The non-negativity of $C$ can thus be relaxed, however, this relaxation is beyond the scope of this paper.
\end{remark}

\section{Conclusion}\label{sec.conclu}

In this paper, we propose a novel model of opinion dynamics in a social network with static topology. Our model is a significant extension of the classical Friedkin-Johnsen model \cite{FriedkinJohnsen:1999} to the case where agents' opinions on two or more interdependent topics are being influenced. The extension is natural if the agent are communicating on several ``logically'' related topics. In the sociological literature, an interdependent set of attitudes and beliefs on multiple
issues is referred to as an ideological or belief system \cite{Converse:1964}. A specification of the interpersonal influence mechanisms and networks that
contribute to the formation of ideological-belief systems has remained an open problem.

We establish necessary and sufficient conditions for the stability of our model and its convergence,
which means that opinions converge to finite limit values for any initial conditions. We also address the problem of identification of the multi-issue interdependence structure.
Although our model requires the agents to communicate synchronously, we show that the same final opinions can be reached by use of a decentralized and asynchronous gossip-based protocol.

Several potential topics of  future research are concerned with experimental validation of our models for large sets of data and investigation of their system-theoretic properties such as e.g. robustness and controllability.
An important open problem is the stability of an extension of the model~\eqref{eq.fjmodel2-2}, where the agents' MiDS matrices are heterogeneous.
A numerical analysis of a system with two different MiDS matrices is available in our recent paper~\cite{FriedkinProTempoPar:2016}, further developing the theory of logically constrained
belief systems formation.

\bibliographystyle{IEEETran}
\bibliography{consensus}

%\begin{thebibliography}{99}
%-------------------------
%\bibitem{FrascaRavazzietal2013} P. Frasca, C. Ravazzi, R. Tempo, H. Ishii
%\newblock Gossips and Prejudices: ergodic randomized dynamics in social networks,
%\newblock \emph{Proc. NecSys-2013}, 2013.

%-------------------------
%\bibitem{FJ1999} N.E.~Friedkin  and E.C.~Johnsen, Social Influence Networks and Opinion Change.
%\newblock \emph{Advances in Group Processes}, No 16,  pp. 1--29, 1999.

%\end{thebibliography}

\appendix
\section*{Stochastic regular matrices}

We state a spectral criterion for regularity.
\begin{lemma}\label{lem.regul}\cite[Ch.XIII, \S7]{GantmacherVol2}.
A row-stochastic square matrix $A$ is regular if and only if $\det(\la I-A)\ne 0$ whenever $\la\ne 1$ and $|\la|=1$; in other words, all eigenvalues of $A$ except for $1$ lie strictly inside
the unit circle. A regular matrix is fully regular if and only if $1$ is a simple eigenvalue, i.e. $1_d$ is the only eigenvector at $1$ up to rescaling: $Az=z\Rightarrow z=c1_d,\,c\in\r$.
\end{lemma}

In the case of irreducible \cite{GantmacherVol2} matrix $A$ regularity and full regularity are both equivalent to the property called \emph{primitivity}, i.e.
strict positivity of the matrix $A^m$ for some $m\ge 0$ which implies that all states of the irreducible Markov chain, generated by $A$, are aperiodic \cite{GantmacherVol2}.
Lemma~\ref{lem.regul} also gives a geometric interpretation of the matrix $A_*$. Let the spectrum of $A$ be $\la_1=1,\la_2,\ldots,\la_d$, where $|\la_j|<1$ as $j>1$.
Then $\r^d$ can be decomposed into a direct sum of invariant \emph{root} subspaces $\r^d=\bigoplus\limits_{j=1}^dL_j$, corresponding to the eigenvalues $\la_j$.
Moreover, the algebraic and geometric multiplicities of $\la_1=1$ always coincide \cite[Ch.XIII,\S6]{GantmacherVol2}, so $L_1$ consists of eigenvectors.
Therefore, the restrictions $A_j=A|_{L_j}$ of $A$ onto $L_j$ are Schur stable for $j>1$, whereas $A_1$ is the identity operator.
Considering a decomposition of an arbitrary vector $v=\sum_jv_j$, where $v_j\in L_j$, one has $A^kv_1=v_1$ and $A^kv_j\to 0$ as $k\to\infty$ for any $j>1$.
Therefore, the operator $A_*:v\mapsto v_1$ is simply the \emph{projector} onto the subspace $L_1$.

As a consequence, we now can easily obtain the equality \eqref{eq.a*}.
Indeed, taking a decomposition $v=v_1+\ldots+v_d$, one easily notices that $(I-\alpha A)^{-1}v_1=(1-\alpha)^{-1}v_1$ and $(I-\alpha A)^{-1}v_i\to (I-A_i)^{-1}v_i$ as $\alpha\to 1$ for any $i>1$.
Hence $\lim\limits_{\alpha\to 1}(I-\alpha A)^{-1}(1-\alpha)v=v_1=A_*v$, which proves \eqref{eq.a*}.

\end{document}